\documentclass[aps,prd,superscriptaddress,nofootinbib,tighten,preprint]{revtex4}
\newcommand{\be}{\begin{eqnarray}}
\newcommand{\ee}{\end{eqnarray}}

\def\nue{{\nu_e}}
\def\anue{{\bar\nu_e}}
\def\numu{{\nu_{\mu}}}
\def\anumu{{\bar\nu_{\mu}}}

\newcommand{\ms}{\Delta m^2_{21}}
\newcommand{\ma}{\Delta m^2_{31}}

\def\gs{\mathrel{
   \rlap{\raise 0.511ex \hbox{$>$}}{\lower 0.511ex \hbox{$\sim$}}}}
\def\ls{\mathrel{
   \rlap{\raise 0.511ex \hbox{$<$}}{\lower 0.511ex \hbox{$\sim$}}}}
\newcommand{\bea}{\begin{equation} \begin{array}{c}}
\newcommand{\bead}{\begin{equation} \begin{array}{cccc}}
\newcommand{\eea}{ \end{array} \end{equation}}
\usepackage{amsfonts}
\usepackage{amssymb}
\usepackage{amsmath}
\usepackage{graphicx}
\usepackage{hyperref}
\usepackage{placeins}
\def\slc#1{\setbox0=\hbox{$#1$}           
    \dimen0=\wd0                                 
    \setbox1=\hbox{/} \dimen1=\wd1               
    \ifdim\dimen0>\dimen1                        
       \rlap{\hbox to \dimen0{\hfil/\hfil}}      
       #1                                        
    \else                                        
       \rlap{\hbox to \dimen1{\hfil$#1$\hfil}}   
       /                                         
    \fi}

\begin{document}

\title{Constraints on Sterile Neutrino Oscillations using DUNE Near Detector}

\author{Sandhya Choubey}
\email{sandhya@hri.res.in}
\affiliation{Harish-Chandra Research Institute, Chhatnag Road, Jhunsi, Allahabad 211 019, India}
\affiliation{Department of Theoretical Physics, School of
Engineering Sciences, KTH Royal Institute of Technology, AlbaNova
University Center, 106 91 Stockholm, Sweden}
\affiliation{Homi Bhabha National Institute, Training School Complex, Anushakti Nagar,
     Mumbai 400085, India
}
\author{Dipyaman Pramanik}
\email{dipyamanpramanik@hri.res.in}
\affiliation{Harish-Chandra Research Institute, Chhatnag Road, Jhunsi, Allahabad 211 019, India}
\affiliation{Homi Bhabha National Institute, Training School Complex, Anushakti Nagar,
     Mumbai 400085, India
}

\begin{abstract}
DUNE (Deep Underground Neutrino Experiment) is a proposed long-baseline neutrino experiment in the US with a baseline of 1300 km from Fermi National Accelerator Laboratory (Fermilab)  to Sanford Underground Research Facility, which will house a 40 kt Liquid Argon Time Projection Chamber (LArTPC) as the far detector. The experiment will also have a fine grained near detector for accurately measuring the initial fluxes. We show that the energy range of the fluxes and baseline of the DUNE near detector is conducive for observing $\nu_\mu \to \nu_e$ oscillations of $\Delta m^2 \sim$ eV$^2$ scale sterile neutrinos, and hence can be effectively used for testing to very high accuracy the reported oscillation signal seen by the LSND and MiniBooNE experiments. We study the sensitivity of the DUNE near detector to sterile neutrino oscillations by varying the baseline, detector fiducial mass and systematic uncertainties. We find that the detector mass and baseline of the currently proposed near detector at DUNE will be able to test the entire LSND parameter region with good precision. The dependence of sensitivity on baseline and detector mass is seen to give interesting results, while dependence on systematic uncertainties is seen to be small. 
\end{abstract}
\maketitle
\section{Introduction}

The existence of neutrino oscillation phenomenon is now well established. In the three-generation paradigm there are six independent parameters, the three mixing angles ($\theta_{12},\theta_{13},\theta_{23}$), two independent mass squared differences ($\Delta m^2_{21}, \Delta m^2_{31} $) and one CP violating phase ($\delta_{CP}$). The first evidence of neutrino oscillations came from the deficit of observed $\nu_e$ solar neutrinos over that predicted by the standard solar model \cite{Reines:1965qd,Cleveland:1998nv,Abdurashitov:2003ew,Hampel:1998xg,Fukuda:2002pe,Fukuda:2001nj}.  This was later independently confirmed by the KamLAND reactor antineutrino experiment \cite{Araki:2004mb}. The combined constraints from the solar and KamLAND data give $\Delta m^2_{21} \simeq 7.5\times 10^{-5}$ eV$^2$ and $\sin^2\theta_{12} \simeq 0.3$ \cite{2014JHEP...11..052G,2016arXiv160107777C}. The atmospheric neutrino experiments \cite{Achar:1965ova,Reines:1965qk,Hirata:1988uy} have observed oscillations of $\numu$ and $\anumu$ \cite{Ashie:2005ik,Adamson:2014vgd,Aartsen:2014yll}, which have been confirmed by the accelerator-based experiments K2K \cite{Ahn:2006zza}, MINOS \cite{Adamson:2014vgd}, T2K \cite{Abe:2015awa} and NO$\nu$A \cite{Adamson:2016xxw}. Together these set of experiments are consistent with neutrino oscillations with mass squared difference $|\Delta m^2_{31}| \simeq 2.5\times 10^{-3} eV^2$, while the best-fit value of $\sin^2\theta_{23}$ is still not firmly determined and changes octant depending on what sign one assumes for $\ma$ \cite{2016arXiv160107777C}. The last mixing angle $\theta_{13}$ is now measured to be $\sin^2\theta_{13} =0.02$ mainly from the reactor data \cite{An:2012eh,Ahn:2012nd,Abe:2011fz}, which is confirmed by the accelerator experiments \cite{Abe:2015awa,2016arXiv160105022A}. Finally, there are some tantalising hints for the CP phase to be close to -90$^\circ$ \cite{2014JHEP...11..052G,2016arXiv160107777C,Abe:2015awa,2016arXiv160105022A}. But definitely data from future experiments such are DUNE would be required to make any definitive statement on this issue, as well as on the issues of sign of $\ma$ and the octant of $\theta_{23}$.  

In addition to the well established oscillations observed in the solar and atmospheric sectors discussed above, signal for neutrino flavor conversion was also reported by the LSND experiment at Los Alamos National Laboratory \cite{2001PhRvD..64k2007A}, which observed a 3.8$\sigma$ excess of $\bar{\nu}_e$ events consistent with $\bar\nu_{\mu}\rightarrow\bar\nu_e$ neutrino oscillations driven by $\Delta m^2 \sim \mathcal{O}$ (eV$^2$). Since oscillations at these frequencies are completely incompatible with those in the solar and atmospheric sectors (whose preferred $\Delta m^2$ are given in the previous paragraph), we need existence of one or more additional neutrino states. Since the number of light neutrinos are constrained to be 3 from the Z invisible decay width measured at LEP \cite{2005hep.ex....9008T}, this implies that the additional neutrino must be sterile. The possible neutrino mass spectra consistent with global data is the so-called 3+1 scenario \cite{1997PhRvD..55.2931G} with one additional sterile neutrino, and 3+2 \cite{2007PhRvD..75a3011K} and 1+3+1 \cite{2006PhRvD..74e3010C} with two sterile neutrinos. The LSND signal was tested at the KARMEN experiment \cite{2002PhRvD..65k2001A} and then at the MiniBooNE experiment \cite{AguilarArevalo:2007it,Aguilar-Arevalo:2013pmq,AguilarArevalo:2010wv}. While KARMEN did not observe any oscillation signal, it failed to rule out the entire LSND allowed parameter region. The MiniBooNE experiment, in its neutrino run, did not observe electron excess in the $L/E$ region where LSND observed an excess. However, they did report an excess of electron events consistent with LSND in their antineutrino channel. In both their neutrino and antineutrino runs, MiniBooNE also saw an excess in the low energy part of the spectrum, which is unlikely to have come from neutrino oscillations. The global fits of these short baseline appearance experiments leave a reasonably large allowed area in the $\Delta m^2-\sin^22\theta_{\mu e}$ plane, where $\Delta m^2$ is the active-sterile mass splitting and $\sin^22\theta_{\mu e}$ is the effective active-sterile mixing angle.
 The major challenge to the sterile neutrino oscillations scenario comes from the short baseline (SBL) $\numu$ disappearance experiments such as CDHS \cite{Dydak:1983zq}, MINOS \cite{Adamson:2010wi}, Super-Kamiokande atmospheric \cite{Abe:2014gda} and MiniBooNE (disappearance search) \cite{Cheng:2012yy}, none of which observed any deficit of $\numu$ and $\anumu$ at all. Therefore global analysis show strong tension between various data sets, allowing the 3+1 case at the 3.7\% C.L.  only \cite{2013JHEP...05..050K,2013ARNPS..63...45C,2013PhRvD..88g3008G}. Addition of the second sterile neutrino brings an improvement in the goodness-of-fit (19\% \cite{2013JHEP...05..050K}), but the tension remains. Hence, more experimental inputs are needed in order to say something conclusively.  Oscillations involving sterile neutrinos have also been proposed as possible explanation of the so-called reactor \cite{2011PhRvC..83e4615M,2011PhRvD..83g3006M,2011PhRvC..84b4617H} and Gallium \cite{1995PhLB..348..121B,2007MPLA...22.2499G,2010PhRvD..82e3005G,2011PhRvC..83f5504G} anomalies, where $\anue$ and $\nue$ disappearance, respectively, have been seen at short baselines. These anomalies also demand a $\Delta m^2 \sim \mathcal{O}$ (eV$^2$), but since they observe disappearance of the first generation of neutrinos, the effective active-sterile mixing angle involved in this case is $\sin^22\theta_{ee}$ \cite{Dewhurst:2015aba,An:2014bik}. While a large number of dedicated experiments are being planned to test these hints of oscillations at such short baselines, in this work we focus on how effectively the DUNE near detector can be used to constrain the active-sterile mass and mixing parameters. 

DUNE (Deep Underground Neutrino Experiment) \cite{Acciarri:2015uup,Acciarri:2016crz,Strait:2016mof,Acciarri:2016ooe} is a proposed long-baseline neutrino experiment in the US with a baseline of 1300 km, where a $\numu$ ($\anumu$) beam will be sent from Fermi National Accelerator Laboratory (Fermilab) in Batavia, Illinois  to Sanford Underground Research Facility in Lead, South Dakota. The far detector at the Sanford Lab will be a 40 kt Liquid Argon Time Projection Chamber (LArTPC), which will observe $\nue$ ($\anue$) appearance and $\numu$ ($\anumu$) disappearance. The main physics goals of DUNE is to determine the 3 unmeasured neutrino oscillation parameters, the mass ordering (sign of $\ma$), CP-violation and the octant of $\theta_{23}$. 
The physics reach of DUNE has been re-evaluated in the presence of non-standard neutrino interactions \cite{Hollander:2014iha,2016arXiv160301380M,2015arXiv151105562D,2016JHEP...03..016C}, sterile neutrinos \cite{2015JHEP...11..039G,2016arXiv160303759A,2015arXiv150703986B} and large extra dimensions \cite{2016arXiv160300018B}. The possibility of constraining active-sterile mixing with the DUNE far detector was considered in \cite{2015arXiv151105562D}. Effect of active-sterile mixing on long-baseline neutrino oscillation experiments have been considered before for MINOS, T2K and NO$\nu$A \cite{Bhattacharya:2011ee,2015PhRvD..91g3017K,2015arXiv150903148P}. Authors of \cite{2014JHEP...12..120B} studied the possibility of testing the LSND parameters space at the near detector of the proposed ESS experiment in Sweden. Here we probe the possibility of confirming or ruling out the LSND parameter space using the DUNE near detector. We check the impact of the baseline, detector fiducial mass and systematic uncertainties on the sensitivity of the DUNE near detector to sterile neutrino oscillations.

The rest of the paper is organised as follows. In section 2 we outline our simulation details. Section 3 gives the main results on this paper. We end with conclusions in section 4. 

\section{The Experimental Set-up}

DUNE \cite{Acciarri:2015uup,Acciarri:2016crz,Strait:2016mof,Acciarri:2016ooe} is an international project, proposed to be built in the US. The experimental set-up consists of $\nu_{\mu}$ ($\bar{\nu}_{\mu}$) beam sent from the Long Baseline Neutrino Facility (LBNF) at Fermilab to the 40 kt of LArTPC far detector at the Sanford Lab in South Dakota at a distance of 1300 km. The LBNF beam is a wide band beam peaked at around 2.5~GeV, giving a $L/E$  which allows for high precision determination of the neutrino oscillation parameters. There is a proposal to also build a high resolution near detector. While the details of the near detector is still under discussion, it is currently proposed to have a fiducial mass of 5-8 t \footnote{In this work we have used a fiducial mass of 5 t for the near detector.}, is expected to be placed at a baseline of 595 m \cite{near,Acciarri:2016ooe} and is designed to measure the initial flux with very high precision. This will help in constraining the systematic uncertainties in the oscillation studies. 
 If there is/are no extra neutrino(s), the near detector will see almost no oscillations as the baseline is very small. However, in presence of extra neutrino states with non-zero mixing, the $\numu \to \nue$ oscillation probability at short distance will be non-zero. If $\Delta m^2\sim\mathcal{O}$(eV$^2$), then the phase $ \Delta m^2 L/4E \sim \pi/2$ for $E\sim\mathcal{O}$(GeV) at small baseline of $\sim \mathcal{O}$(km). Since the near detector will be placed at a baseline of this order, we expect the DUNE near detector to observe full oscillations of the sterile neutrinos, if indeed the LSND claim is correct. 

\begin{table}[]
\centering
\begin{tabular}{|l|l|}
\hline
Baseline & 595 m /1km /3km \\ \hline
Fiducial mass & 5t /400t /1kt \\ \hline
Detector type & HiResMnu \\ \hline
Energy Resolution $e$ & $6\%/\sqrt{E}$\\ \hline
Energy Resolution $\mu$ & $3.7\%$ \\ \hline
Signal Normalization Error & 1\%  \\ \hline
Background Normalization Error & 5\%  \\ \hline
Energy Calibration Error & 2\%  \\ \hline
Energy Range & 0-8 GeV \\ \hline
Bin Width & 0.25 GeV \\ \hline
Backgrounds & 0.1\% $\nu_{\mu}$  CC mis-identification , 0.1\% NC background \\ \hline

\end{tabular}
\caption{\label{tab:near} The DUNE near detector specification used in this work.}
\end{table}
\FloatBarrier

We next give the simulation details used for generating the results in this work. We have used GLoBES (Global Long Baseline Experiment Simulator) \cite{2005CoPhC.167..195H,2007CoPhC.177..432H} for generating the numerical results in our analysis of the DUNE experiment. The proposed near detector HiResMnu has an active tracker of dimension $350\times350\times750$ cm$^3$, surrounded by an ECAL embedded in a dipole magnet with $B\sim0.4 T$. The fiducial volume corresponds to $\sim5t$ of mass. The active target is composed of straw tube trackers. The detector shall have Argon target at the upstream end of the tracker \cite{near}. The benchmark near detector configuration in our analysis is summarised in the Table \ref{tab:near}. The energy resolution, detector mass and baseline are taken from the near detector specifications given in the DUNE near detector document \cite{near}. In addition to the benchmark choices for these parameters given in \cite{near}, we use two other choices of baseline and detector mass to show the impact of these on the sensitivity of the experiment. For detector mass, we also show results for 400 t  and 1 kt. The mass of 400 t is motivated by the ProtoDUNE proposal, which is expected to have 400 t of LArTPC \cite{Cavanna:2144868} and could at some point be placed along the LBNF beam-line. The 1 kt mass is just an ad-hoc choice of a very large fiducial mass to show that the experiment reaches its statistical saturation already at 400 t and the LSND sensitivity becomes insensitive to any further increase in statistics. For baseline, we also show results for 1 km and 3 km. For the systematic uncertainties, since the possible near detector systematics are not yet clear, we assume the same as that given for the DUNE far detector \cite{2013arXiv1307.7335L}. We vary this benchmark systematic uncertainties to showcase their impact of the mass and mixing sensitivity. 

We use the DUNE flux provided by \cite{marie} for this work. A beam power of 1.2 MW and an exposure of 5 years in $\numu$ and 5 years in $\anumu$ mode is used in our analysis. We use the GLoBES package \cite{2005CoPhC.167..195H,2007CoPhC.177..432H} for simulating the DUNE experiment. The neutrino oscillation probabilities in the presence of sterile neutrino are calculated using snu, which is a neutrino oscillation code for calculating oscillation probabilities for GLoBES in the presence of sterile neutrino \cite{Kopp:2006wp,Kopp:2007ne}.

\section{Results}

As at short baselines, the oscillations due to $\ma$ and $\ms$ do not develop, at such distance we can approximate the oscillation probability  by an effective two generation framework, which can be written as

\begin{equation}
P(\nu_{\mu}\rightarrow\nu_e) \simeq \sin^22\theta_{\mu e}\sin^2(\frac{\Delta m^2L}{4E})
\,
\label{5}
\end{equation}
where $\theta_{\mu e}$ is the effective mixing angle and $\Delta m^2$ is the new mass squared difference. Throughout this article we work in this approximation and study everything in terms of electron appearance data only.

\begin{figure}[]
\begin{center}
\includegraphics[width=0.45\textwidth]{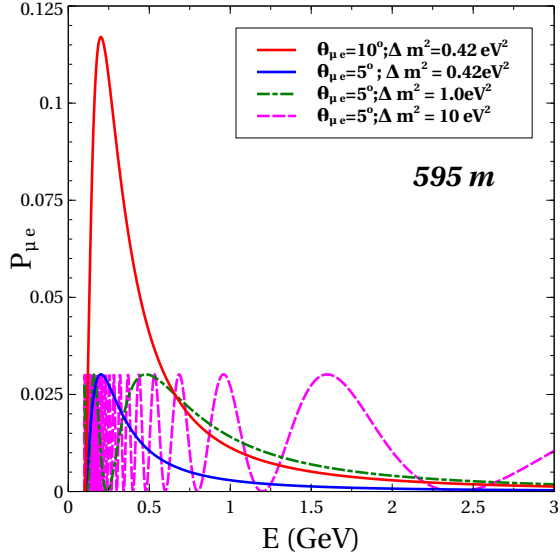}
\caption{ $P_{\mu e}$ probability at 595 m for different sterile neutrino oscilation parameters. 
}
\label{prob}
\end{center}
\end{figure}

The  Fig.~\ref{prob} shows the probability of $\nu_{\mu}$ going to $\nu_e$ as a function of energy 
 at the 595 m baseline, in presence of one extra sterile neutrino. The red solid curve depicts probability for $\theta_{\mu e} = 10^\circ$ and $\Delta m^2=0.42$ eV$^2$. The blue solid, green dot-dashed and magenta dashed curves are for $\theta_{\mu e}=5^\circ$ and $\Delta m^2=0.42$ eV$^2$, 1.0 eV$^2$ and 10.0 eV$^2$, respectively. 
 We see that for higher mixing angles the amplitude of the oscillation increases, and for higher mass-squared differences the oscillations become faster. In all of the cases depicted in Fig. \ref{prob}, we can see substantial oscillations between $\nu_{\mu}$ and $\nu_e$ in presence of sterile neutrino. The DUNE beam has huge flux in this energy window. Therefore, DUNE near detector should be highly sensitive in this region of the parameter space.

\begin{figure}[t!]
\begin{center}
\includegraphics[width=0.45\textwidth]{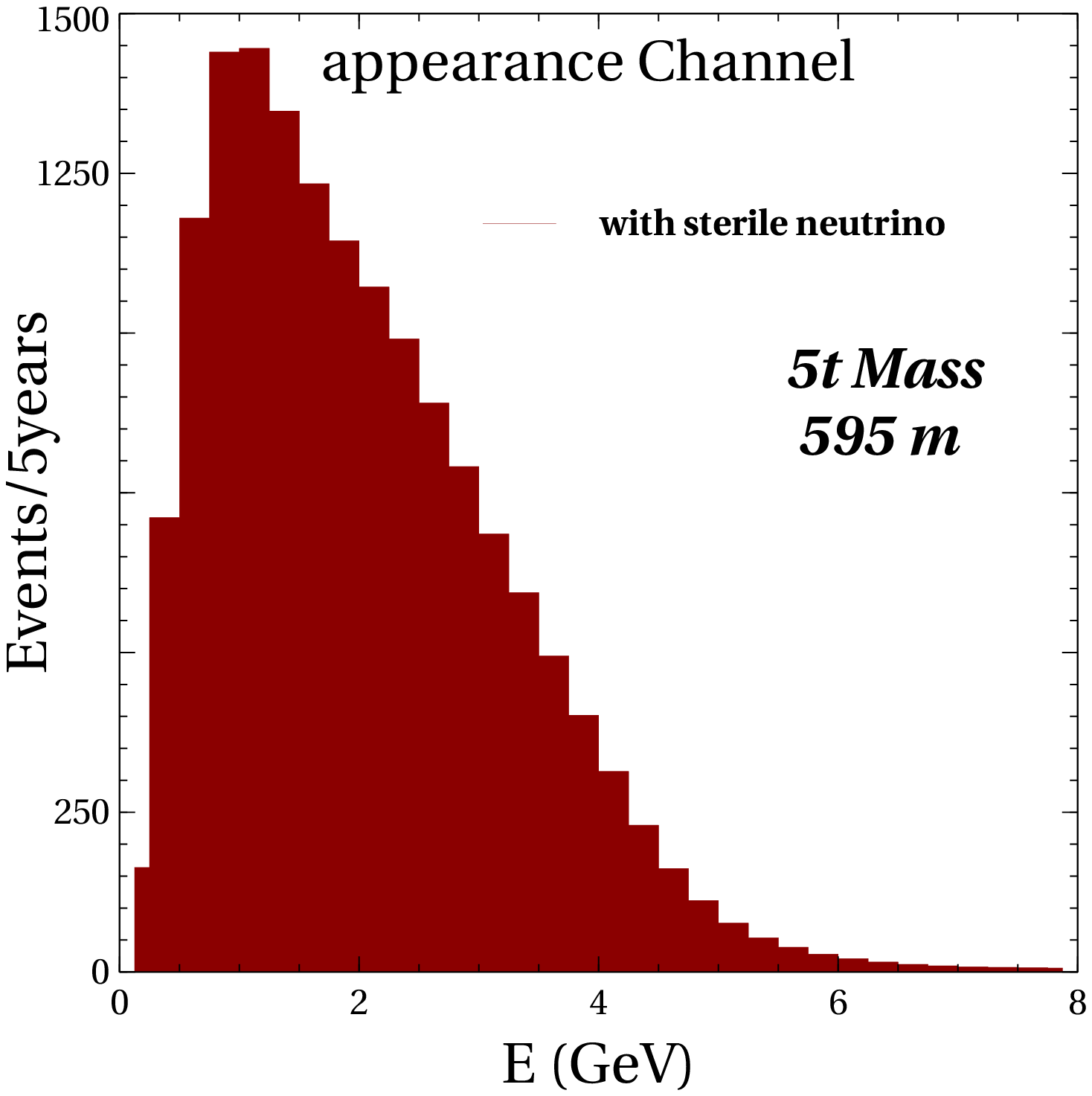}
\includegraphics[width=0.45\textwidth]{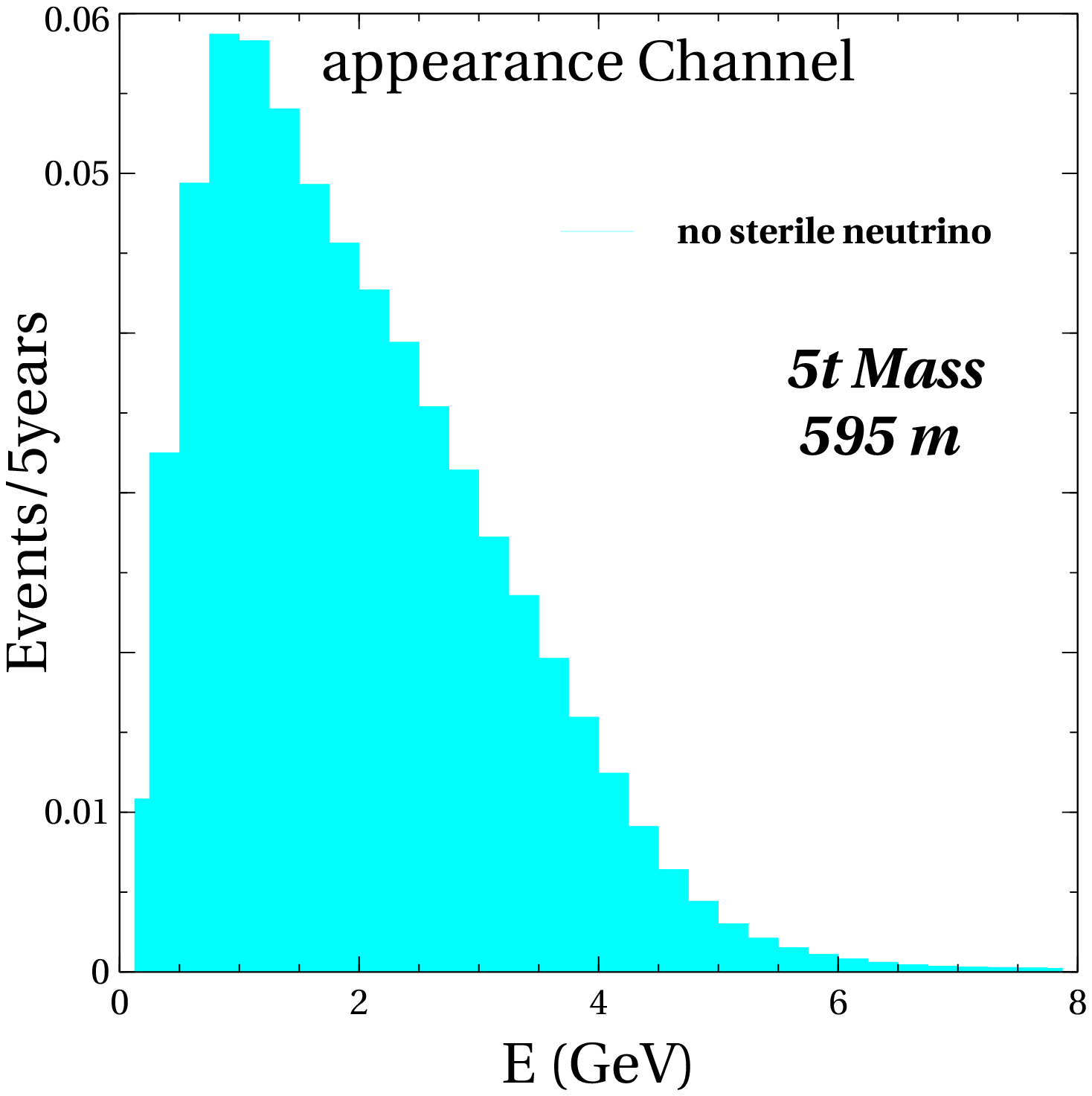}

\caption{Left: 5 years of events at neutrino appearance channel for the case with sterile neutrino at a 5t near detector kept at a baseline of 595 m. Right: 5 years of events at neutrino appearance channel for the case with no sterile neutrino with same detector configuration.}
\label{eventap}
\end{center}
\end{figure}

Fig.~\ref{eventap} shows the event rates for 5 years exposure from a 1.2 MW beam at the near detector for $\theta_{\mu e}=5^\circ$ and $\Delta m^2=0.42$ eV$^2$ for the appearance channel. The left panel shows the event rate when there is a sterile neutrino, whereas the right panel shows the event rate when there is no sterile neutrino.We can see that oscillations due to the sterile neutrino state changes the number of events significantly for the appearance channel. Note that the high flux at the near detector magnifies the effects of any oscillation at this small baseline. 

We next define a $\chi^2$ as \cite{2013arXiv1307.7335L},
\begin{equation}
\chi^2(\textbf{n}^{true},\textbf{n}^{test},f) = 2 \sum_{i}^{N_{reco}}\bigg (n_i^{true}\ln\frac{n_i^{true}}{n_i^{test}(f)}+n_i^{test}(f)-n_i^{true}\bigg )+f^2
\,,
\label{5}
\end{equation} 
where $n_i$ is the number of events, $i$ is the bin index, $f$ represents the nuisance parameters, and $'true'$ and $'test'$ represent data and fit respectively. The systematic uncertainties and backgrounds assumed in the simulations are given in Table \ref{tab:near}. Using Eq. (\ref{5}) we present exclusion plots in the $\Delta m^2-\sin^22\theta_{\mu e}$ plane. In order to do that we take no sterile neutrino in data and scan the  $\Delta m^2-\sin^22\theta_{\mu e}$ parameter space in the fit. The following plots depict how much region of the LSND  parameter space can be excluded by the DUNE near detector. In our fit we consider 
only the appearance data. 

%

\begin{figure}[]
\begin{center}
\vskip 0.8cm
\includegraphics[width=0.45\textwidth]{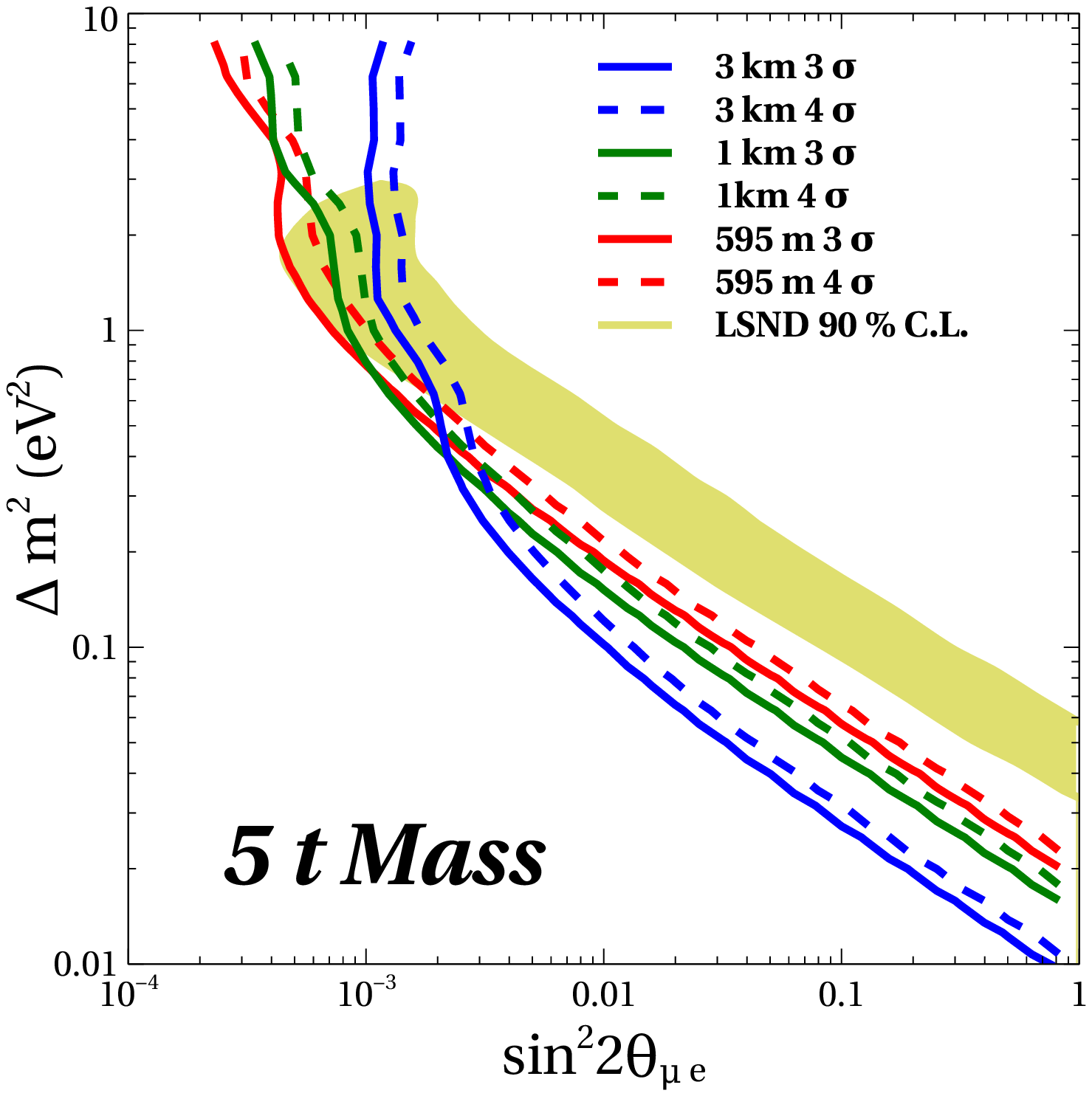}
\includegraphics[width=0.45\textwidth]{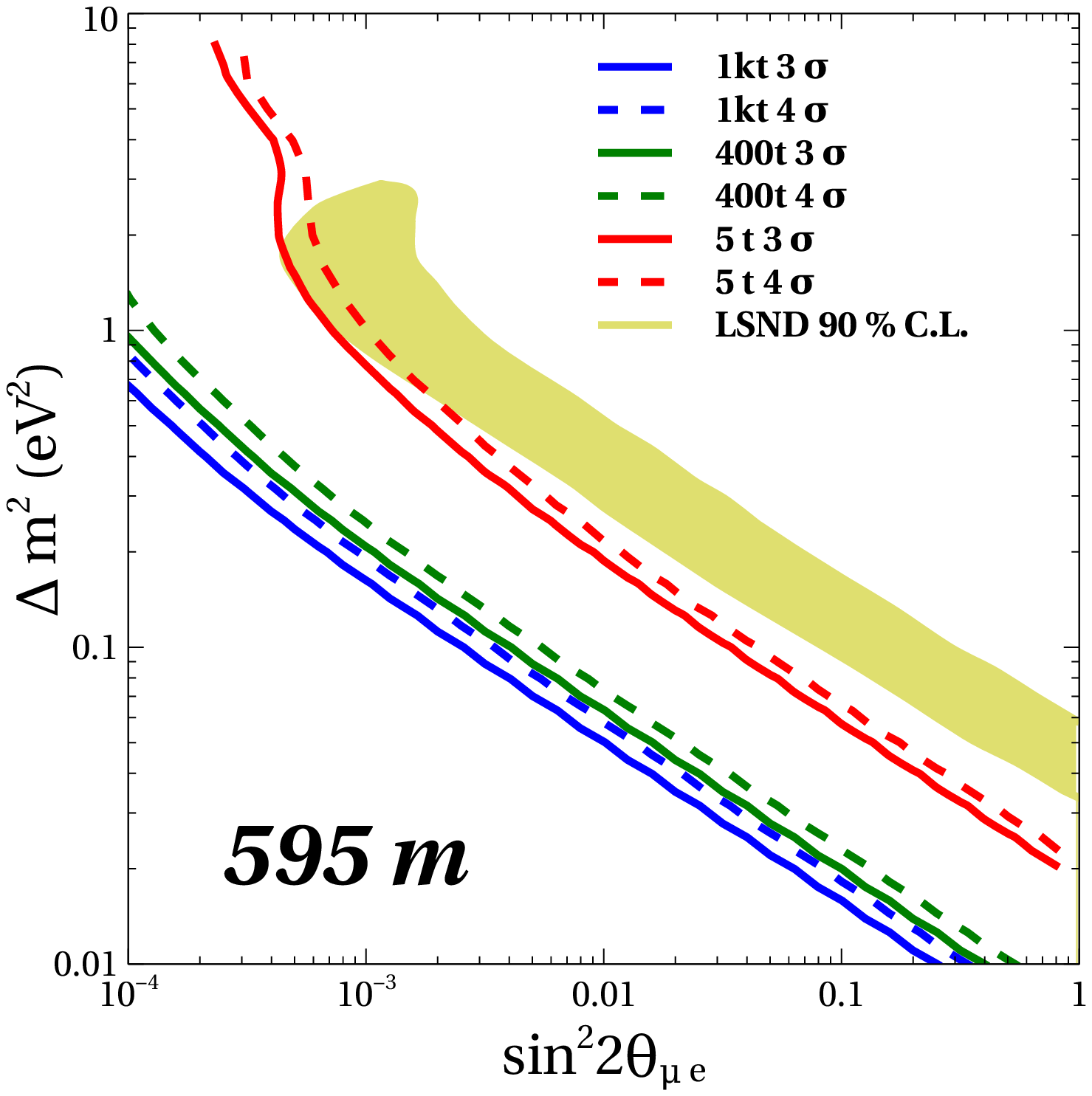}
\caption{Left: Exclusion contours at 3 \& 4 $\sigma$ confidence levels for near detector of mass 5t with different baselines. Right: Exclusion contours at 3 \& 4 $\sigma$ confidence levels for near detector at 595 m baseline with different detector mass. We consider an exposure of 5+5 for $\numu + \anumu$. }
\label{chi2}
\end{center}
\end{figure}

\begin{figure}[]
\begin{center}
\includegraphics[width=0.45\textwidth]{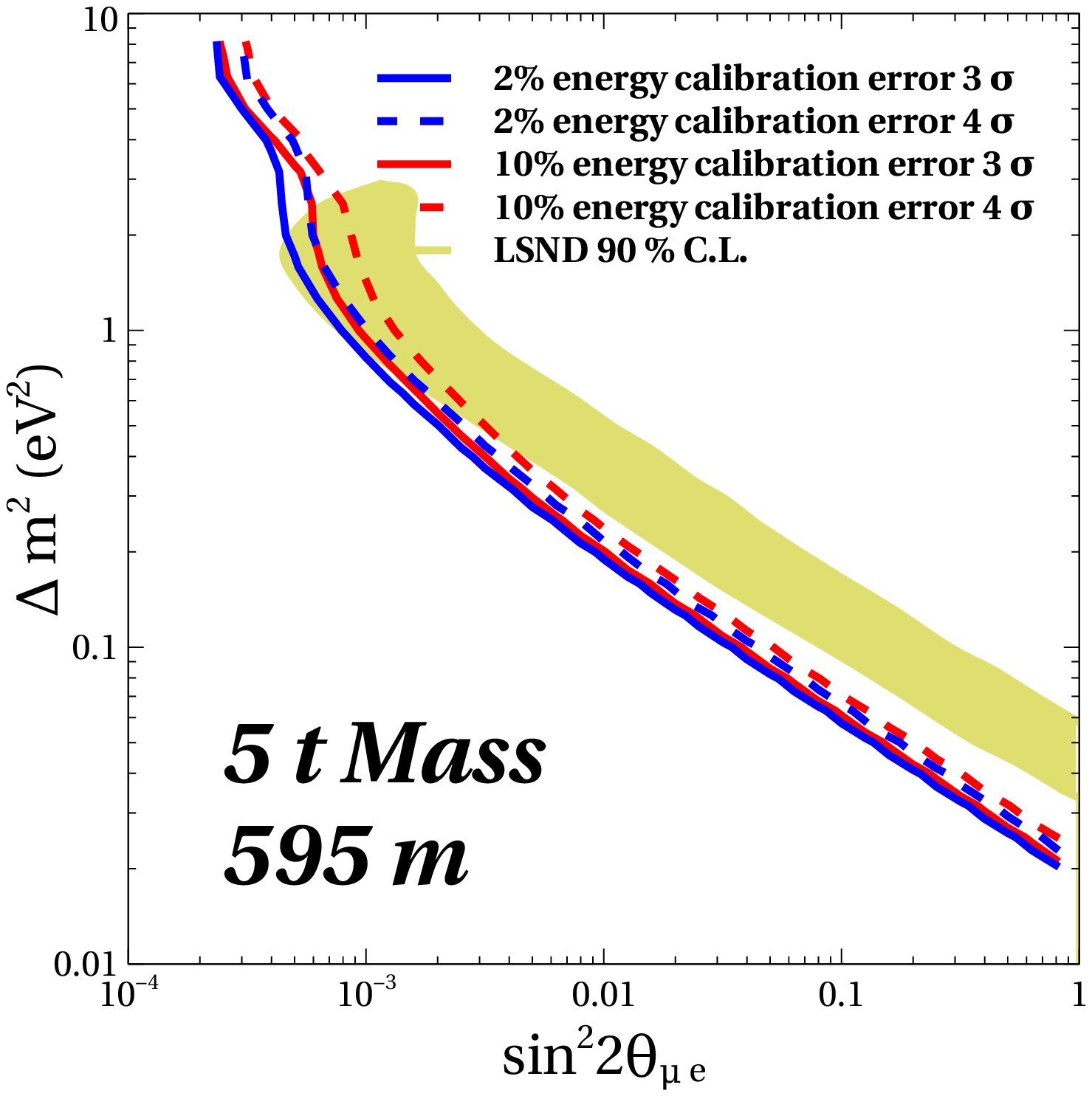}
\includegraphics[width=0.45\textwidth]{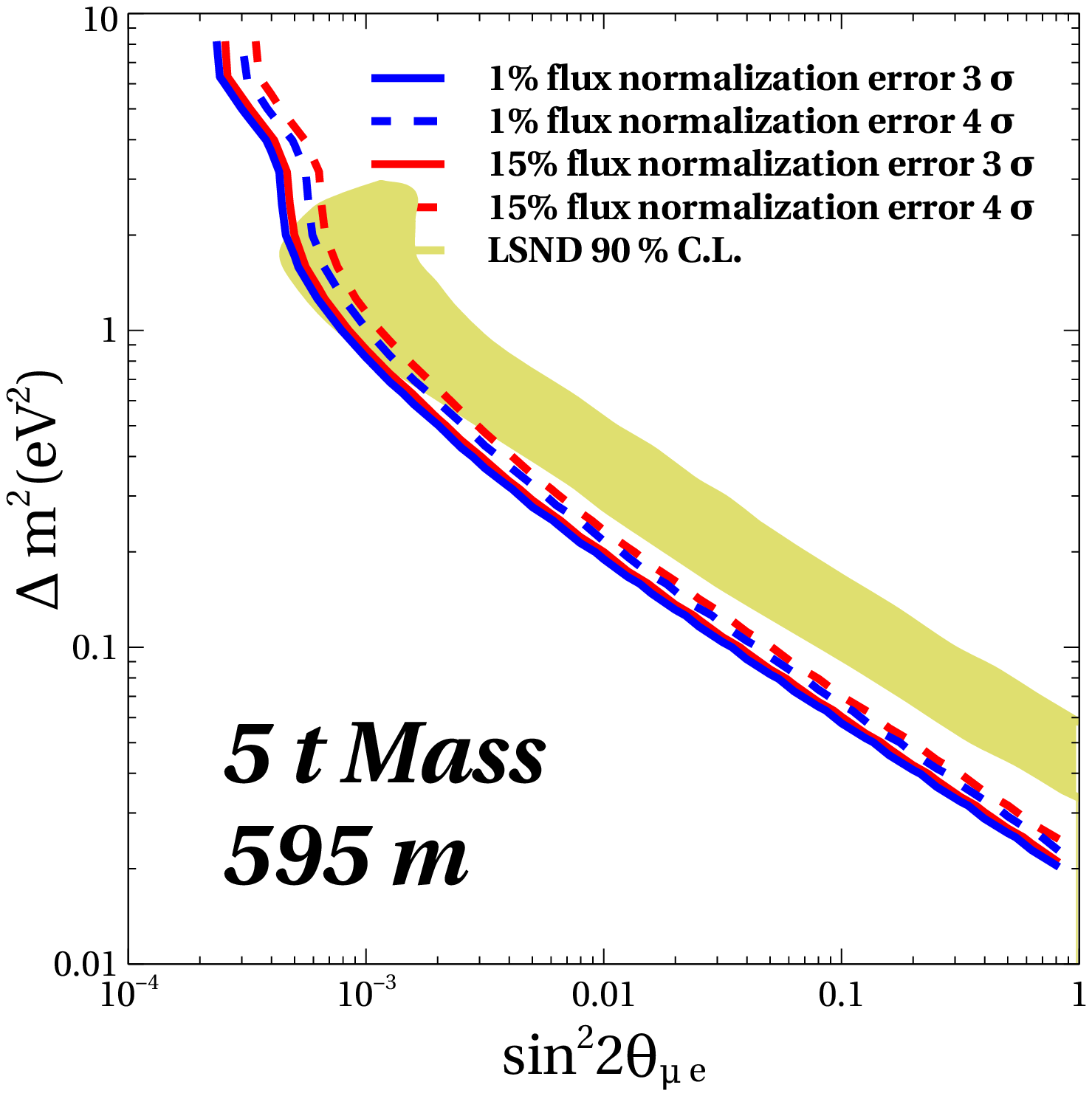}

\caption{Left:Exclusion contours for at 3 \& 4 $\sigma$ confidence levels for different energy calibration errors for DUNE near detector at 595 m baseline with 5t mass. Right: Exclusion contours for 3 \& and 4 $\sigma$ confidence levels for different signal normalisation error. We consider an exposure of 5+5 for $\numu + \anumu$. }
\label{sys}
\end{center}
\end{figure}

The plots in Fig.~\ref{chi2} are exclusion curves for different near detector configurations. The simulation is done for 5+5 years of $\nu_{\mu}$, $\bar{\nu}_{\mu}$ beam at 1.2 MW. The solid and dashed lines give the $3\sigma$ and $4\sigma$ exclusion limits respectively. The yellow shaded area shows the 90\% C.L. allowed region from LSND \cite{2001PhRvD..64k2007A}. The left panel shows how exclusion limit changes as we go to higher baselines, for a fixed detector mass of 5 t. We have kept the other detector parameters same as described in Table \ref{tab:near}. We can see that the DUNE near detector is able to almost rule out the LSND results \cite{2001PhRvD..64k2007A} for all the baseline options. However, we notice that the curve showing exclusion limit changes shape as we change the baseline. We see that as we move to longer baselines, the sensitivity to lower mixing angle is reduced while sensitivity to mass-squared difference at higher mixing angle is increased. The main reason is that the statistics at the higher baselines are $1/L^2$ suppressed, leading to loss of sensitivity for lower mixing angles for which the oscillation probability is proportionally suppressed. However, longer baselines allow for oscillations of lower $\Delta m^2$ better, leading for better sensitivity for these parameter regions. The shorter baseline on the other hand has higher statistics due to lower $L$, allowing it to measure lower mixing angles better, however, the oscillations for lower $\Delta m^2$ do not develop and the corresponding sensitivity drops. \footnote{The decay pipe of the DUNE beam has a length of about 210 m. This could bring an uncertainty in the distance of flight of the neutrinos which might have a bearing on the oscillation signal at the near detector. We have explicitly checked that the impact of this uncertainty is not very significant. We stress that the sensitivity plots shown in this paper are for illustration only.}

The right panel of Fig.~\ref{chi2} gives the variation of the exclusion limits as we vary the fiducial mass. We have kept the baseline fixed at 595 m for all cases in this panel. Results for the three benchmark masses of 5 t, 400 t and 1 kt are shown. The 5 t is chosen for it is given in the DUNE near detector proposal \cite{near}, 400 t is chosen because it is going to be the mass of the ProtoDUNE detector \cite{Cavanna:2144868} and 1 kt is just another benchmark point. We can see that the 400 t configuration can comfortably rule out the LSND result. The 5 t detector can also almost rule out the LSND allowed region with about $3\sigma$ C.L.. The figure also shows that fiducial mass has reached its plateau at 400 t, such that any further increase in detector mass and/or exposure does not change the sensitivity by any significant amount. So 400 t LArTPC ProtoDUNE placed at a baseline of about 595 m  can be a good choice for the near detector for testing LSND. We have checked that the detector energy resolution does not bring any significant change  to our final results.

\begin{figure}[]
\begin{center}
\includegraphics[width=0.45\textwidth]{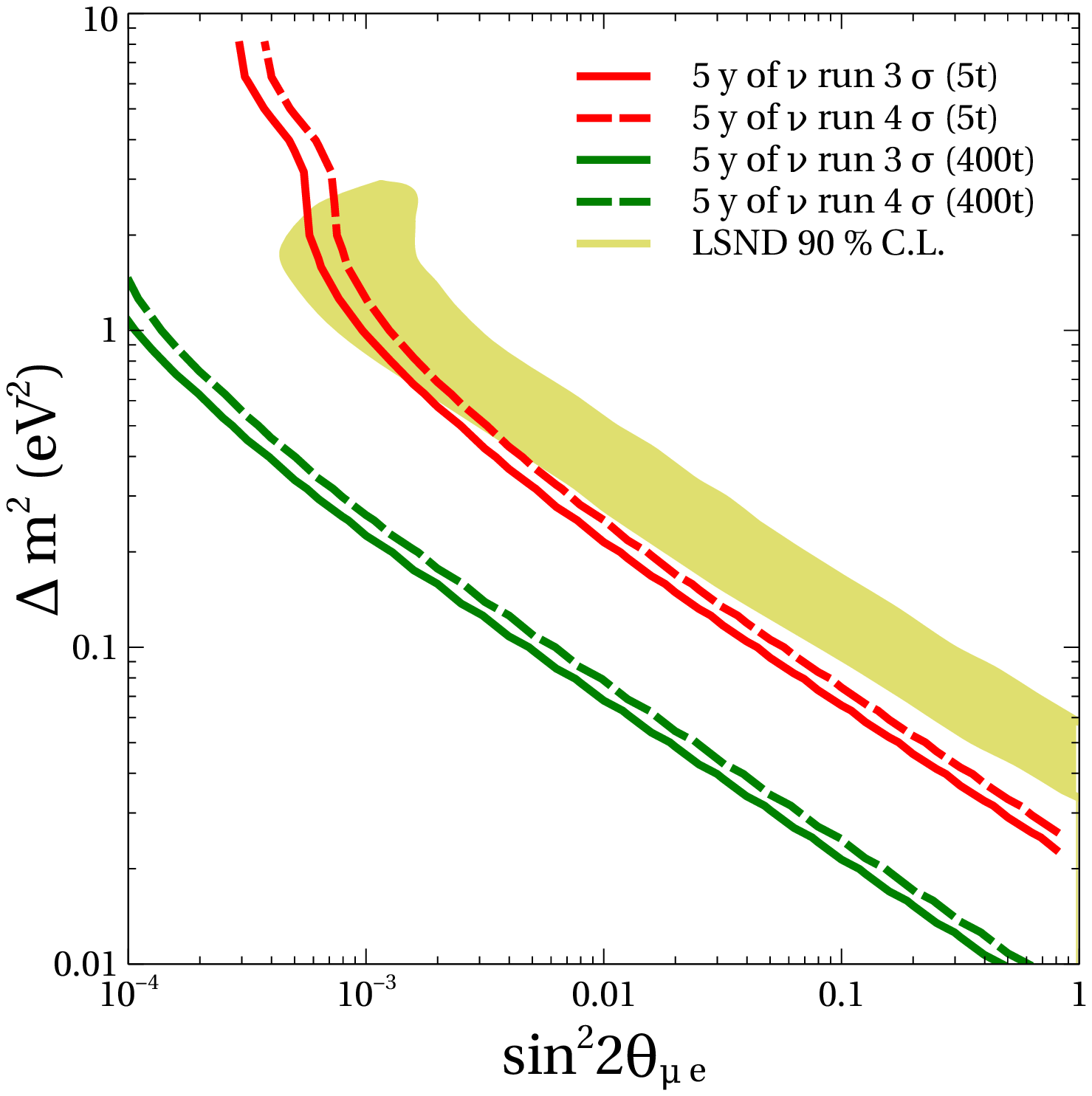}
\includegraphics[width=0.45\textwidth]{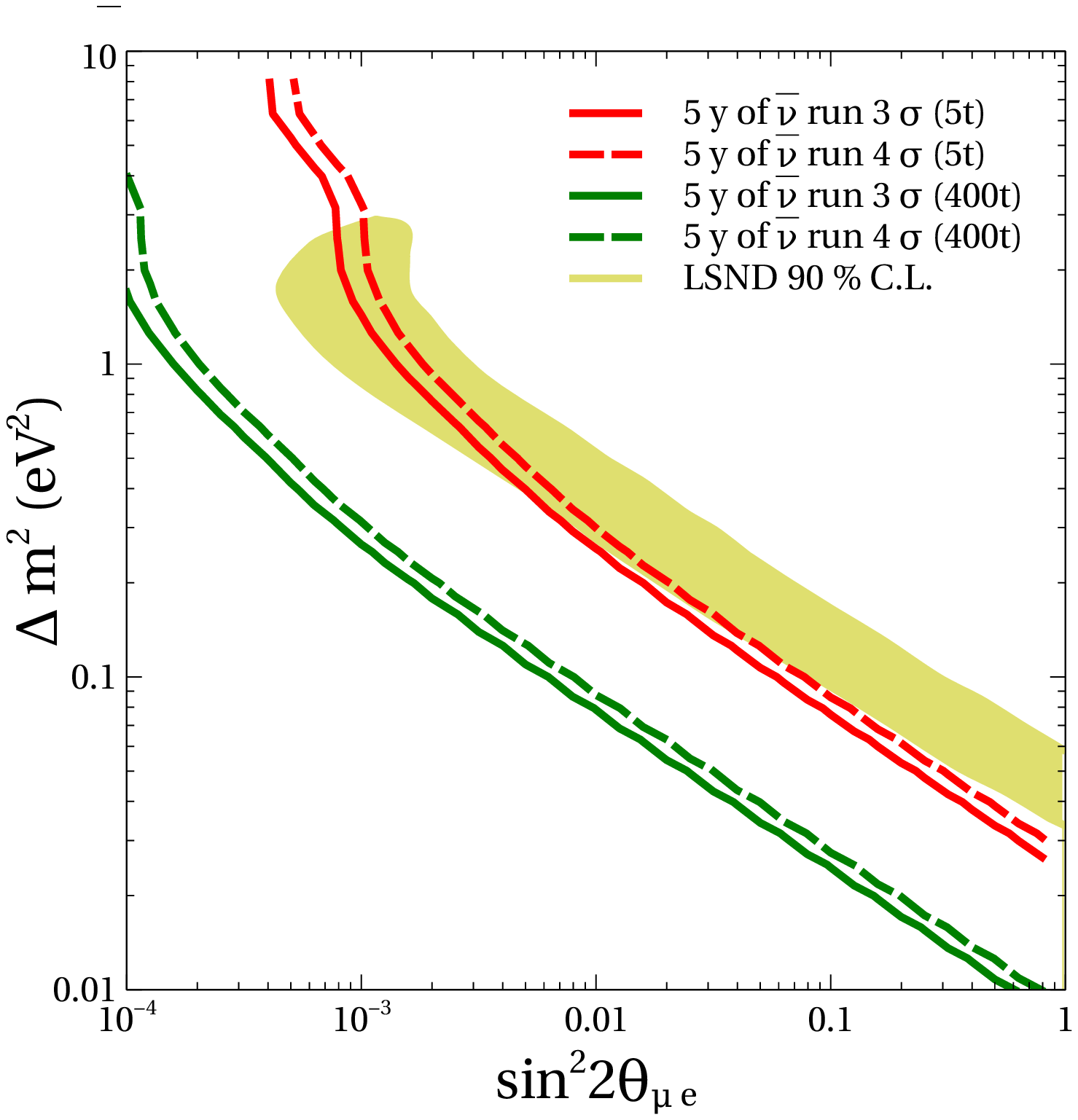}
\caption{\label{fig:nuanu}The left (right) panel shows the result for running DUNE for 5 years in the neutrino (antineutrino) mode. In both panels, the solid lines are the exclusion curves for $3 \sigma$ C.L. and the dashed lines are the exclusion curves for $4 \sigma $ C.L. The red lines are for 5 t detector and green lines are for 400 t detectors, while the yellow shaded region is the 90\% C.L. LSND allowed region.}
\end{center}
\end{figure}


In Fig.~\ref{sys} we have shown the effect of systematic uncertainties on the sensitivity of the near detector to the LSND region. We have considered one optimal (2\%) and one conservative (10\%) energy calibration error as systematic error in the left panel. In the right panel we have considered 1\% and 15\% flux normalisation error of signal. We can see from this figure that the systematic uncertainties have very small effect on the sensitivity.

In Fig. \ref{fig:nuanu} we study the sensitivity of DUNE near detectors separately in the neutrino and the antineutrino modes. Both LSND  \cite{2001PhRvD..64k2007A} and MiniBooNE \cite{AguilarArevalo:2007it,Aguilar-Arevalo:2013pmq,AguilarArevalo:2010wv} have reported electron excess compatible with oscillations only in the antineutrino channel. Since DUNE will be running for 5 years in the neutrino and 5 years in the antineutrino mode, we can probe sterile neutrino oscillations separately in the two data sets. The results of our study has been shown in Fig. \ref{fig:nuanu}. The left (right) panel shows the result for running DUNE for 5 years in the neutrino (antineutrino) mode. In both cases we show the sensitivity curves for the 5 t and 400 t detector mass cases. One can see that the neutrino mode can better exclude the parameter space for both 5 t as well as 400 t options, but the anti-neutrino mode also performs well and can test the LSND region.

\begin{figure}[]
\begin{center}
\includegraphics[width=0.5\textwidth]{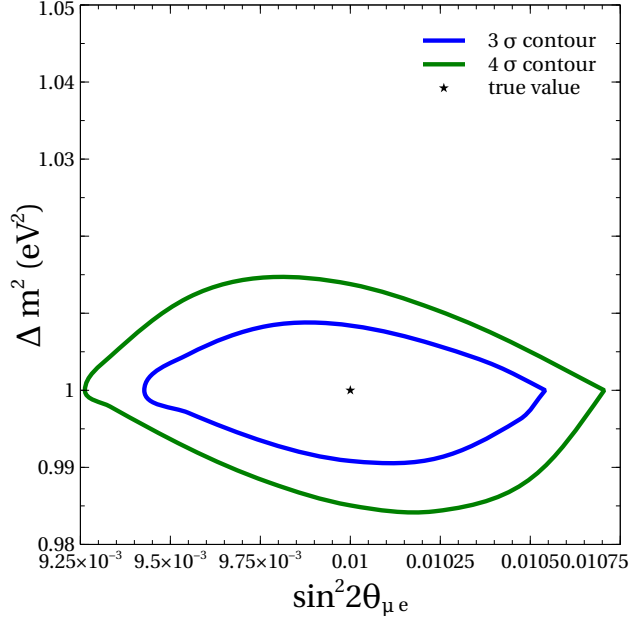}
\caption{\label{fig:cont}Expected sensitivity at the $3 \sigma$ and $4 \sigma$ C.L. for the case when there is a sterile neutrino in the data. The black star is the value for which the data is generated. The blue contour shows the $3 \sigma$ and the green contour shows the $4 \sigma$ C.L.}
\end{center}
\end{figure} 

There are quite a few experimental proposals that plan to test the LSND signal before DUNE near detector becomes operational. If any of these experiments manage to see a signal compatible with sterile neutrino oscillations, then it is pertinent to ask how well the DUNE near detector could measure the $\Delta m^2$ and $\sin^22\theta_{\mu e}$. To that end we perform a study of the projected reach of the DUNE near detector by generating the data at one benchmark point in the $\Delta m^2 - \sin^22\theta_{\mu e}$ plane. This data is then fitted and the results shown in Fig.~\ref{fig:cont}. The black star shows the benchmark point chosen. The blue contour shows the bound at $3 \sigma$ C.L. and the green contour shows the $4 \sigma$ C.L. We see  that the sensitivity is very good at the $3 \sigma$ and $4 \sigma$ confidence levels and DUNE near detector can tightly constrain the parameter space.

\section{Summary \& Conclusion}

In this letter we have discussed the reach of the DUNE near detector in constraining the LSND parameter space. Since the oscillations due to $\ms$ and $\ma$ are irrelevant at such short baselines, we presented our results in the effective two-generation $\Delta m^2-\sin^22\theta_{\mu e}$ parameter space. We showed that for the DUNE near detector baseline of 595 m, the entire LSND allowed region can be probed well above the $4 \sigma $ C.L.. Since the near detector configuration is not yet completely final, we studied the sensitivity of the experiment to test the LSND region as a function of the detector mass, the detector baseline as well as the detector systematics. We concluded that the 595 m baseline is most optimal for this study and a mass of 400 t, as proposed for the ProtoDUNE detector, would give very sensitive results though even the proposed HiResMnu with 5 t mass can almost rule out the LSND parameters space. We also showed that the detector systematic uncertainties do not bring any significant difference to the sensitivity of the near detector to short baseline neutrino oscillations.

In conclusion, while the main physics goal of the 
DUNE experimental proposal is the measurement of CP phase $\delta_{CP}$, we 
showed in this paper that very good sensitivity to the LSND parameter space comes as 
a bonus from the data at the near detector of this experiment. 
A large number of experiments have been proposed to test the LSND claim. These include experiments like Short Baseline Neutrino Program(SBN \cite{Antonello:2015lea}), stopped pion beam experiments (OscSNS \cite{OscSNS:2013hua}, LSND reloaded \cite{Agarwalla:2010zu}),
kaon decay at rest beams \cite{Spitz:2012gp}
and decay in flight neutrino beams (BooNE, LArTPC detectors at CERN \cite{Antonello:2012hf}, 
MicroBooNE \cite{Jones:2011ci}, Very Low Energy Neutrino Factory \cite{Koshkarev:1974my,Neuffer:1980ru}). 
All of these experiments are expected to test the entire allowed LSND region within a 
relatively short time-frame. In this letter we looked at the expected sensitivity of the 
DUNE experiment to the LSND parameter space. 
A comparison of the DUNE sensitivity to those 
expected from the dedicated experiments proposed to test LSND reveals that for the 
400 t detector mass case (protoDUNE mass), the expected sensitivity of DUNE is comparable,
if not better, than some of these proposals. Even for the 5 t mass case, 
the sensitivity is rather good. 
If any of the future experiments (mentioned in \cite{Antonello:2015lea,OscSNS:2013hua,Agarwalla:2010zu,Spitz:2012gp,Antonello:2012hf,Jones:2011ci,Koshkarev:1974my}) is able to find sterile neutrino oscillations, we showed that the DUNE near detector is a nice option to constrain the parameter space further. 
Therefore, the DUNE experiment provides an independent probe of the 
LSND anomaly at no extra cost since the experiment, including the near 
detector, will be built (if funded) for a different and very important physics reason, 
the study of CP violation in the lepton sector. 

\section*{Acknowledgment}
The authors would thank the Neutrino Project under the XII plan of Harish-Chandra Research Institute. 
This project has received funding from the European Union's Horizon 2020 research and innovation programme under the Marie Sklodowska-Curie grant agreement No 674896. And this project has received funding from the European Union’s Horizon 2020 research and innovation programme under the Marie Skłodowska-Curie grant agreement No 690575.
DP also acknowledges Mary Bishai for providing the DUNE flux and M. Masud for the help in simulation using GLoBES.

\bibliography{ref}

\begin{thebibliography}{79}
\expandafter\ifx\csname natexlab\endcsname\relax\def\natexlab#1{#1}\fi
\expandafter\ifx\csname bibnamefont\endcsname\relax
  \def\bibnamefont#1{#1}\fi
\expandafter\ifx\csname bibfnamefont\endcsname\relax
  \def\bibfnamefont#1{#1}\fi
\expandafter\ifx\csname citenamefont\endcsname\relax
  \def\citenamefont#1{#1}\fi
\expandafter\ifx\csname url\endcsname\relax
  \def\url#1{\texttt{#1}}\fi
\expandafter\ifx\csname urlprefix\endcsname\relax\def\urlprefix{URL }\fi
\providecommand{\bibinfo}[2]{#2}
\providecommand{\eprint}[2][]{\url{#2}}

\bibitem[{\citenamefont{Reines and Woods}(1965)}]{Reines:1965qd}
\bibinfo{author}{\bibfnamefont{F.}~\bibnamefont{Reines}} \bibnamefont{and}
  \bibinfo{author}{\bibfnamefont{R.~M.} \bibnamefont{Woods}},
  \bibinfo{journal}{Phys. Rev. Lett.} \textbf{\bibinfo{volume}{14}},
  \bibinfo{pages}{20} (\bibinfo{year}{1965}).

\bibitem[{\citenamefont{Cleveland et~al.}(1998)\citenamefont{Cleveland, Daily,
  Davis, Distel, Lande, Lee, Wildenhain, and Ullman}}]{Cleveland:1998nv}
\bibinfo{author}{\bibfnamefont{B.~T.} \bibnamefont{Cleveland}},
  \bibinfo{author}{\bibfnamefont{T.}~\bibnamefont{Daily}},
  \bibinfo{author}{\bibfnamefont{R.}~\bibnamefont{Davis}, \bibfnamefont{Jr.}},
  \bibinfo{author}{\bibfnamefont{J.~R.} \bibnamefont{Distel}},
  \bibinfo{author}{\bibfnamefont{K.}~\bibnamefont{Lande}},
  \bibinfo{author}{\bibfnamefont{C.~K.} \bibnamefont{Lee}},
  \bibinfo{author}{\bibfnamefont{P.~S.} \bibnamefont{Wildenhain}},
  \bibnamefont{and} \bibinfo{author}{\bibfnamefont{J.}~\bibnamefont{Ullman}},
  \bibinfo{journal}{Astrophys. J.} \textbf{\bibinfo{volume}{496}},
  \bibinfo{pages}{505} (\bibinfo{year}{1998}).

\bibitem[{\citenamefont{Abdurashitov et~al.}(2003)}]{Abdurashitov:2003ew}
\bibinfo{author}{\bibfnamefont{J.~N.} \bibnamefont{Abdurashitov}}
  \bibnamefont{et~al.} (\bibinfo{collaboration}{SAGE}), \bibinfo{journal}{Nucl.
  Phys. Proc. Suppl.} \textbf{\bibinfo{volume}{118}}, \bibinfo{pages}{39}
  (\bibinfo{year}{2003}), \bibinfo{note}{[,39(2003)]}.

\bibitem[{\citenamefont{Hampel et~al.}(1999)}]{Hampel:1998xg}
\bibinfo{author}{\bibfnamefont{W.}~\bibnamefont{Hampel}} \bibnamefont{et~al.}
  (\bibinfo{collaboration}{GALLEX}), \bibinfo{journal}{Phys. Lett.}
  \textbf{\bibinfo{volume}{B447}}, \bibinfo{pages}{127} (\bibinfo{year}{1999}).

\bibitem[{\citenamefont{Fukuda et~al.}(2002)}]{Fukuda:2002pe}
\bibinfo{author}{\bibfnamefont{S.}~\bibnamefont{Fukuda}} \bibnamefont{et~al.}
  (\bibinfo{collaboration}{Super-Kamiokande}), \bibinfo{journal}{Phys. Lett.}
  \textbf{\bibinfo{volume}{B539}}, \bibinfo{pages}{179} (\bibinfo{year}{2002}),
  \eprint{hep-ex/0205075}.

\bibitem[{\citenamefont{Fukuda et~al.}(2001)}]{Fukuda:2001nj}
\bibinfo{author}{\bibfnamefont{S.}~\bibnamefont{Fukuda}} \bibnamefont{et~al.}
  (\bibinfo{collaboration}{Super-Kamiokande}), \bibinfo{journal}{Phys. Rev.
  Lett.} \textbf{\bibinfo{volume}{86}}, \bibinfo{pages}{5651}
  (\bibinfo{year}{2001}), \eprint{hep-ex/0103032}.

\bibitem[{\citenamefont{Araki et~al.}(2005)}]{Araki:2004mb}
\bibinfo{author}{\bibfnamefont{T.}~\bibnamefont{Araki}} \bibnamefont{et~al.}
  (\bibinfo{collaboration}{KamLAND}), \bibinfo{journal}{Phys. Rev. Lett.}
  \textbf{\bibinfo{volume}{94}}, \bibinfo{pages}{081801}
  (\bibinfo{year}{2005}), \eprint{hep-ex/0406035}.

\bibitem[{\citenamefont{{Gonzalez-Garcia}
  et~al.}(2014)\citenamefont{{Gonzalez-Garcia}, {Maltoni}, and
  {Schwetz}}}]{2014JHEP...11..052G}
\bibinfo{author}{\bibfnamefont{M.~C.} \bibnamefont{{Gonzalez-Garcia}}},
  \bibinfo{author}{\bibfnamefont{M.}~\bibnamefont{{Maltoni}}},
  \bibnamefont{and}
  \bibinfo{author}{\bibfnamefont{T.}~\bibnamefont{{Schwetz}}},
  \bibinfo{journal}{Journal of High Energy Physics}
  \textbf{\bibinfo{volume}{11}}, \bibinfo{eid}{52} (\bibinfo{year}{2014}),
  \eprint{1409.5439}.

\bibitem[{\citenamefont{{Capozzi} et~al.}(2016)\citenamefont{{Capozzi}, {Lisi},
  {Marrone}, {Montanino}, and {Palazzo}}}]{2016arXiv160107777C}
\bibinfo{author}{\bibfnamefont{F.}~\bibnamefont{{Capozzi}}},
  \bibinfo{author}{\bibfnamefont{E.}~\bibnamefont{{Lisi}}},
  \bibinfo{author}{\bibfnamefont{A.}~\bibnamefont{{Marrone}}},
  \bibinfo{author}{\bibfnamefont{D.}~\bibnamefont{{Montanino}}},
  \bibnamefont{and}
  \bibinfo{author}{\bibfnamefont{A.}~\bibnamefont{{Palazzo}}},
  \bibinfo{journal}{ArXiv e-prints}  (\bibinfo{year}{2016}),
  \eprint{1601.07777}.

\bibitem[{\citenamefont{Achar et~al.}(1965)}]{Achar:1965ova}
\bibinfo{author}{\bibfnamefont{C.~V.} \bibnamefont{Achar}}
  \bibnamefont{et~al.}, \bibinfo{journal}{Phys. Lett.}
  \textbf{\bibinfo{volume}{18}}, \bibinfo{pages}{196} (\bibinfo{year}{1965}).

\bibitem[{\citenamefont{Reines et~al.}(1965)\citenamefont{Reines, Crouch,
  Jenkins, Kropp, Gurr, Smith, Sellschop, and Meyer}}]{Reines:1965qk}
\bibinfo{author}{\bibfnamefont{F.}~\bibnamefont{Reines}},
  \bibinfo{author}{\bibfnamefont{M.~F.} \bibnamefont{Crouch}},
  \bibinfo{author}{\bibfnamefont{T.~L.} \bibnamefont{Jenkins}},
  \bibinfo{author}{\bibfnamefont{W.~R.} \bibnamefont{Kropp}},
  \bibinfo{author}{\bibfnamefont{H.~S.} \bibnamefont{Gurr}},
  \bibinfo{author}{\bibfnamefont{G.~R.} \bibnamefont{Smith}},
  \bibinfo{author}{\bibfnamefont{J.~P.~F.} \bibnamefont{Sellschop}},
  \bibnamefont{and} \bibinfo{author}{\bibfnamefont{B.}~\bibnamefont{Meyer}},
  \bibinfo{journal}{Phys. Rev. Lett.} \textbf{\bibinfo{volume}{15}},
  \bibinfo{pages}{429} (\bibinfo{year}{1965}).

\bibitem[{\citenamefont{Hirata et~al.}(1988)}]{Hirata:1988uy}
\bibinfo{author}{\bibfnamefont{K.~S.} \bibnamefont{Hirata}}
  \bibnamefont{et~al.} (\bibinfo{collaboration}{Kamiokande-II}),
  \bibinfo{journal}{Phys. Lett.} \textbf{\bibinfo{volume}{B205}},
  \bibinfo{pages}{416} (\bibinfo{year}{1988}), \bibinfo{note}{[,447(1988)]}.

\bibitem[{\citenamefont{Ashie et~al.}(2005)}]{Ashie:2005ik}
\bibinfo{author}{\bibfnamefont{Y.}~\bibnamefont{Ashie}} \bibnamefont{et~al.}
  (\bibinfo{collaboration}{Super-Kamiokande}), \bibinfo{journal}{Phys. Rev.}
  \textbf{\bibinfo{volume}{D71}}, \bibinfo{pages}{112005}
  (\bibinfo{year}{2005}), \eprint{hep-ex/0501064}.

\bibitem[{\citenamefont{Adamson et~al.}(2014)}]{Adamson:2014vgd}
\bibinfo{author}{\bibfnamefont{P.}~\bibnamefont{Adamson}} \bibnamefont{et~al.}
  (\bibinfo{collaboration}{MINOS}), \bibinfo{journal}{Phys. Rev. Lett.}
  \textbf{\bibinfo{volume}{112}}, \bibinfo{pages}{191801}
  (\bibinfo{year}{2014}), \eprint{1403.0867}.

\bibitem[{\citenamefont{Aartsen et~al.}(2015)}]{Aartsen:2014yll}
\bibinfo{author}{\bibfnamefont{M.}~\bibnamefont{Aartsen}} \bibnamefont{et~al.}
  (\bibinfo{collaboration}{IceCube}), \bibinfo{journal}{Phys. Rev.}
  \textbf{\bibinfo{volume}{D91}}, \bibinfo{pages}{072004}
  (\bibinfo{year}{2015}), \eprint{1410.7227}.

\bibitem[{\citenamefont{Ahn et~al.}(2006)}]{Ahn:2006zza}
\bibinfo{author}{\bibfnamefont{M.~H.} \bibnamefont{Ahn}} \bibnamefont{et~al.}
  (\bibinfo{collaboration}{K2K}), \bibinfo{journal}{Phys. Rev.}
  \textbf{\bibinfo{volume}{D74}}, \bibinfo{pages}{072003}
  (\bibinfo{year}{2006}), \eprint{hep-ex/0606032}.

\bibitem[{\citenamefont{Abe et~al.}(2015{\natexlab{a}})}]{Abe:2015awa}
\bibinfo{author}{\bibfnamefont{K.}~\bibnamefont{Abe}} \bibnamefont{et~al.}
  (\bibinfo{collaboration}{T2K}), \bibinfo{journal}{Phys. Rev.}
  \textbf{\bibinfo{volume}{D91}}, \bibinfo{pages}{072010}
  (\bibinfo{year}{2015}{\natexlab{a}}), \eprint{1502.01550}.

\bibitem[{\citenamefont{Adamson et~al.}(2016)}]{Adamson:2016xxw}
\bibinfo{author}{\bibfnamefont{P.}~\bibnamefont{Adamson}} \bibnamefont{et~al.}
  (\bibinfo{collaboration}{NOvA}), \bibinfo{journal}{Phys. Rev.}
  \textbf{\bibinfo{volume}{D93}}, \bibinfo{pages}{051104}
  (\bibinfo{year}{2016}), \eprint{1601.05037}.

\bibitem[{\citenamefont{An et~al.}(2012)}]{An:2012eh}
\bibinfo{author}{\bibfnamefont{F.~P.} \bibnamefont{An}} \bibnamefont{et~al.}
  (\bibinfo{collaboration}{Daya Bay}), \bibinfo{journal}{Phys. Rev. Lett.}
  \textbf{\bibinfo{volume}{108}}, \bibinfo{pages}{171803}
  (\bibinfo{year}{2012}), \eprint{1203.1669}.

\bibitem[{\citenamefont{Ahn et~al.}(2012)}]{Ahn:2012nd}
\bibinfo{author}{\bibfnamefont{J.~K.} \bibnamefont{Ahn}} \bibnamefont{et~al.}
  (\bibinfo{collaboration}{RENO}), \bibinfo{journal}{Phys. Rev. Lett.}
  \textbf{\bibinfo{volume}{108}}, \bibinfo{pages}{191802}
  (\bibinfo{year}{2012}), \eprint{1204.0626}.

\bibitem[{\citenamefont{Abe et~al.}(2012)}]{Abe:2011fz}
\bibinfo{author}{\bibfnamefont{Y.}~\bibnamefont{Abe}} \bibnamefont{et~al.}
  (\bibinfo{collaboration}{Double Chooz}), \bibinfo{journal}{Phys. Rev. Lett.}
  \textbf{\bibinfo{volume}{108}}, \bibinfo{pages}{131801}
  (\bibinfo{year}{2012}), \eprint{1112.6353}.

\bibitem[{\citenamefont{{Adamson} et~al.}(2016)\citenamefont{{Adamson}, {Ader},
  {Andrews}, {Anfimov}, {Anghel}, {Arms}, {Arrieta-Diaz}, {Aurisano}, {Ayres},
  {Backhouse} et~al.}}]{2016arXiv160105022A}
\bibinfo{author}{\bibfnamefont{P.}~\bibnamefont{{Adamson}}},
  \bibinfo{author}{\bibfnamefont{C.}~\bibnamefont{{Ader}}},
  \bibinfo{author}{\bibfnamefont{M.}~\bibnamefont{{Andrews}}},
  \bibinfo{author}{\bibfnamefont{N.}~\bibnamefont{{Anfimov}}},
  \bibinfo{author}{\bibfnamefont{I.}~\bibnamefont{{Anghel}}},
  \bibinfo{author}{\bibfnamefont{K.}~\bibnamefont{{Arms}}},
  \bibinfo{author}{\bibfnamefont{E.}~\bibnamefont{{Arrieta-Diaz}}},
  \bibinfo{author}{\bibfnamefont{A.}~\bibnamefont{{Aurisano}}},
  \bibinfo{author}{\bibfnamefont{D.}~\bibnamefont{{Ayres}}},
  \bibinfo{author}{\bibfnamefont{C.}~\bibnamefont{{Backhouse}}},
  \bibnamefont{et~al.}, \bibinfo{journal}{ArXiv e-prints}
  (\bibinfo{year}{2016}), \eprint{1601.05022}.

\bibitem[{\citenamefont{{Aguilar} et~al.}(2001)\citenamefont{{Aguilar},
  {Auerbach}, {Burman}, {Caldwell}, {Church}, {Cochran}, {Donahue}, {Fazely},
  {Garvey}, {Gunasingha} et~al.}}]{2001PhRvD..64k2007A}
\bibinfo{author}{\bibfnamefont{A.}~\bibnamefont{{Aguilar}}},
  \bibinfo{author}{\bibfnamefont{L.~B.} \bibnamefont{{Auerbach}}},
  \bibinfo{author}{\bibfnamefont{R.~L.} \bibnamefont{{Burman}}},
  \bibinfo{author}{\bibfnamefont{D.~O.} \bibnamefont{{Caldwell}}},
  \bibinfo{author}{\bibfnamefont{E.~D.} \bibnamefont{{Church}}},
  \bibinfo{author}{\bibfnamefont{A.~K.} \bibnamefont{{Cochran}}},
  \bibinfo{author}{\bibfnamefont{J.~B.} \bibnamefont{{Donahue}}},
  \bibinfo{author}{\bibfnamefont{A.}~\bibnamefont{{Fazely}}},
  \bibinfo{author}{\bibfnamefont{G.~T.} \bibnamefont{{Garvey}}},
  \bibinfo{author}{\bibfnamefont{R.~M.} \bibnamefont{{Gunasingha}}},
  \bibnamefont{et~al.}, \bibinfo{journal}{\prd} \textbf{\bibinfo{volume}{64}},
  \bibinfo{pages}{112007} (\bibinfo{year}{2001}), \eprint{hep-ex/0104049}.

\bibitem[{\citenamefont{{The ALEPH Collaboration}
  et~al.}(2005)\citenamefont{{The ALEPH Collaboration}, {the DELPHI
  Collaboration}, {the L3 Collaboration}, {the OPAL Collaboration}, {the SLD
  Collaboration}, {the LEP Electroweak Working Group}, {SLD electroweak}, and
  {heavy flavour groups}}}]{2005hep.ex....9008T}
\bibinfo{author}{\bibnamefont{{The ALEPH Collaboration}}},
  \bibinfo{author}{\bibnamefont{{the DELPHI Collaboration}}},
  \bibinfo{author}{\bibnamefont{{the L3 Collaboration}}},
  \bibinfo{author}{\bibnamefont{{the OPAL Collaboration}}},
  \bibinfo{author}{\bibnamefont{{the SLD Collaboration}}},
  \bibinfo{author}{\bibnamefont{{the LEP Electroweak Working Group}}},
  \bibinfo{author}{\bibfnamefont{t.}~\bibnamefont{{SLD electroweak}}},
  \bibnamefont{and} \bibinfo{author}{\bibnamefont{{heavy flavour groups}}},
  \bibinfo{journal}{ArXiv High Energy Physics - Experiment e-prints}
  (\bibinfo{year}{2005}), \eprint{hep-ex/0509008}.

\bibitem[{\citenamefont{{Goswami}}(1997)}]{1997PhRvD..55.2931G}
\bibinfo{author}{\bibfnamefont{S.}~\bibnamefont{{Goswami}}},
  \bibinfo{journal}{\prd} \textbf{\bibinfo{volume}{55}}, \bibinfo{pages}{2931}
  (\bibinfo{year}{1997}), \eprint{hep-ph/9507212}.

\bibitem[{\citenamefont{{Karagiorgi} et~al.}(2007)\citenamefont{{Karagiorgi},
  {Aguilar-Arevalo}, {Conrad}, {Shaevitz}, {Whisnant}, {Sorel}, and
  {Barger}}}]{2007PhRvD..75a3011K}
\bibinfo{author}{\bibfnamefont{G.}~\bibnamefont{{Karagiorgi}}},
  \bibinfo{author}{\bibfnamefont{A.}~\bibnamefont{{Aguilar-Arevalo}}},
  \bibinfo{author}{\bibfnamefont{J.~M.} \bibnamefont{{Conrad}}},
  \bibinfo{author}{\bibfnamefont{M.~H.} \bibnamefont{{Shaevitz}}},
  \bibinfo{author}{\bibfnamefont{K.}~\bibnamefont{{Whisnant}}},
  \bibinfo{author}{\bibfnamefont{M.}~\bibnamefont{{Sorel}}}, \bibnamefont{and}
  \bibinfo{author}{\bibfnamefont{V.}~\bibnamefont{{Barger}}},
  \bibinfo{journal}{\prd} \textbf{\bibinfo{volume}{75}}, \bibinfo{eid}{013011}
  (\bibinfo{year}{2007}), \eprint{hep-ph/0609177}.

\bibitem[{\citenamefont{{Choubey} et~al.}(2006)\citenamefont{{Choubey},
  {Harries}, and {Ross}}}]{2006PhRvD..74e3010C}
\bibinfo{author}{\bibfnamefont{S.}~\bibnamefont{{Choubey}}},
  \bibinfo{author}{\bibfnamefont{N.~P.} \bibnamefont{{Harries}}},
  \bibnamefont{and} \bibinfo{author}{\bibfnamefont{G.~G.}
  \bibnamefont{{Ross}}}, \bibinfo{journal}{\prd} \textbf{\bibinfo{volume}{74}},
  \bibinfo{eid}{053010} (\bibinfo{year}{2006}), \eprint{hep-ph/0605255}.

\bibitem[{\citenamefont{{Armbruster} et~al.}(2002)\citenamefont{{Armbruster},
  {Blair}, {Bodmann}, {Booth}, {Drexlin}, {Edgington}, {Eichner}, {Eitel},
  {Finckh}, {Gemmeke} et~al.}}]{2002PhRvD..65k2001A}
\bibinfo{author}{\bibfnamefont{B.}~\bibnamefont{{Armbruster}}},
  \bibinfo{author}{\bibfnamefont{I.~M.} \bibnamefont{{Blair}}},
  \bibinfo{author}{\bibfnamefont{B.~A.} \bibnamefont{{Bodmann}}},
  \bibinfo{author}{\bibfnamefont{N.~E.} \bibnamefont{{Booth}}},
  \bibinfo{author}{\bibfnamefont{G.}~\bibnamefont{{Drexlin}}},
  \bibinfo{author}{\bibfnamefont{J.~A.} \bibnamefont{{Edgington}}},
  \bibinfo{author}{\bibfnamefont{C.}~\bibnamefont{{Eichner}}},
  \bibinfo{author}{\bibfnamefont{K.}~\bibnamefont{{Eitel}}},
  \bibinfo{author}{\bibfnamefont{E.}~\bibnamefont{{Finckh}}},
  \bibinfo{author}{\bibfnamefont{H.}~\bibnamefont{{Gemmeke}}},
  \bibnamefont{et~al.}, \bibinfo{journal}{\prd} \textbf{\bibinfo{volume}{65}},
  \bibinfo{eid}{112001} (\bibinfo{year}{2002}), \eprint{hep-ex/0203021}.

\bibitem[{\citenamefont{Aguilar-Arevalo et~al.}(2007)}]{AguilarArevalo:2007it}
\bibinfo{author}{\bibfnamefont{A.~A.} \bibnamefont{Aguilar-Arevalo}}
  \bibnamefont{et~al.} (\bibinfo{collaboration}{MiniBooNE}),
  \bibinfo{journal}{Phys. Rev. Lett.} \textbf{\bibinfo{volume}{98}},
  \bibinfo{pages}{231801} (\bibinfo{year}{2007}), \eprint{0704.1500}.

\bibitem[{\citenamefont{Aguilar-Arevalo
  et~al.}(2013)}]{Aguilar-Arevalo:2013pmq}
\bibinfo{author}{\bibfnamefont{A.~A.} \bibnamefont{Aguilar-Arevalo}}
  \bibnamefont{et~al.} (\bibinfo{collaboration}{MiniBooNE}),
  \bibinfo{journal}{Phys. Rev. Lett.} \textbf{\bibinfo{volume}{110}},
  \bibinfo{pages}{161801} (\bibinfo{year}{2013}), \eprint{1207.4809}.

\bibitem[{\citenamefont{Aguilar-Arevalo et~al.}(2010)}]{AguilarArevalo:2010wv}
\bibinfo{author}{\bibfnamefont{A.~A.} \bibnamefont{Aguilar-Arevalo}}
  \bibnamefont{et~al.} (\bibinfo{collaboration}{MiniBooNE}),
  \bibinfo{journal}{Phys. Rev. Lett.} \textbf{\bibinfo{volume}{105}},
  \bibinfo{pages}{181801} (\bibinfo{year}{2010}), \eprint{1007.1150}.

\bibitem[{\citenamefont{Dydak et~al.}(1984)}]{Dydak:1983zq}
\bibinfo{author}{\bibfnamefont{F.}~\bibnamefont{Dydak}} \bibnamefont{et~al.},
  \bibinfo{journal}{Phys. Lett.} \textbf{\bibinfo{volume}{B134}},
  \bibinfo{pages}{281} (\bibinfo{year}{1984}).

\bibitem[{\citenamefont{Adamson et~al.}(2010)}]{Adamson:2010wi}
\bibinfo{author}{\bibfnamefont{P.}~\bibnamefont{Adamson}} \bibnamefont{et~al.}
  (\bibinfo{collaboration}{MINOS}), \bibinfo{journal}{Phys. Rev.}
  \textbf{\bibinfo{volume}{D81}}, \bibinfo{pages}{052004}
  (\bibinfo{year}{2010}), \eprint{1001.0336}.

\bibitem[{\citenamefont{Abe et~al.}(2015{\natexlab{b}})}]{Abe:2014gda}
\bibinfo{author}{\bibfnamefont{K.}~\bibnamefont{Abe}} \bibnamefont{et~al.}
  (\bibinfo{collaboration}{Super-Kamiokande}), \bibinfo{journal}{Phys. Rev.}
  \textbf{\bibinfo{volume}{D91}}, \bibinfo{pages}{052019}
  (\bibinfo{year}{2015}{\natexlab{b}}), \eprint{1410.2008}.

\bibitem[{\citenamefont{Cheng et~al.}(2012)}]{Cheng:2012yy}
\bibinfo{author}{\bibfnamefont{G.}~\bibnamefont{Cheng}} \bibnamefont{et~al.}
  (\bibinfo{collaboration}{SciBooNE, MiniBooNE}), \bibinfo{journal}{Phys. Rev.}
  \textbf{\bibinfo{volume}{D86}}, \bibinfo{pages}{052009}
  (\bibinfo{year}{2012}), \eprint{1208.0322}.

\bibitem[{\citenamefont{{Kopp} et~al.}(2013)\citenamefont{{Kopp}, {Machado},
  {Maltoni}, and {Schwetz}}}]{2013JHEP...05..050K}
\bibinfo{author}{\bibfnamefont{J.}~\bibnamefont{{Kopp}}},
  \bibinfo{author}{\bibfnamefont{P.~A.~N.} \bibnamefont{{Machado}}},
  \bibinfo{author}{\bibfnamefont{M.}~\bibnamefont{{Maltoni}}},
  \bibnamefont{and}
  \bibinfo{author}{\bibfnamefont{T.}~\bibnamefont{{Schwetz}}},
  \bibinfo{journal}{Journal of High Energy Physics}
  \textbf{\bibinfo{volume}{5}}, \bibinfo{eid}{50} (\bibinfo{year}{2013}),
  \eprint{1303.3011}.

\bibitem[{\citenamefont{{Conrad} et~al.}(2013)\citenamefont{{Conrad}, {Louis},
  and {Shaevitz}}}]{2013ARNPS..63...45C}
\bibinfo{author}{\bibfnamefont{J.~M.} \bibnamefont{{Conrad}}},
  \bibinfo{author}{\bibfnamefont{W.~C.} \bibnamefont{{Louis}}},
  \bibnamefont{and} \bibinfo{author}{\bibfnamefont{M.~H.}
  \bibnamefont{{Shaevitz}}}, \bibinfo{journal}{Annual Review of Nuclear and
  Particle Science} \textbf{\bibinfo{volume}{63}}, \bibinfo{pages}{45}
  (\bibinfo{year}{2013}), \eprint{1306.6494}.

\bibitem[{\citenamefont{{Giunti} et~al.}(2013)\citenamefont{{Giunti},
  {Laveder}, {Li}, and {Long}}}]{2013PhRvD..88g3008G}
\bibinfo{author}{\bibfnamefont{C.}~\bibnamefont{{Giunti}}},
  \bibinfo{author}{\bibfnamefont{M.}~\bibnamefont{{Laveder}}},
  \bibinfo{author}{\bibfnamefont{Y.~F.} \bibnamefont{{Li}}}, \bibnamefont{and}
  \bibinfo{author}{\bibfnamefont{H.~W.} \bibnamefont{{Long}}},
  \bibinfo{journal}{\prd} \textbf{\bibinfo{volume}{88}}, \bibinfo{eid}{073008}
  (\bibinfo{year}{2013}), \eprint{1308.5288}.

\bibitem[{\citenamefont{{Mueller} et~al.}(2011)\citenamefont{{Mueller},
  {Lhuillier}, {Fallot}, {Letourneau}, {Cormon}, {Fechner}, {Giot}, {Lasserre},
  {Martino}, {Mention} et~al.}}]{2011PhRvC..83e4615M}
\bibinfo{author}{\bibfnamefont{T.~A.} \bibnamefont{{Mueller}}},
  \bibinfo{author}{\bibfnamefont{D.}~\bibnamefont{{Lhuillier}}},
  \bibinfo{author}{\bibfnamefont{M.}~\bibnamefont{{Fallot}}},
  \bibinfo{author}{\bibfnamefont{A.}~\bibnamefont{{Letourneau}}},
  \bibinfo{author}{\bibfnamefont{S.}~\bibnamefont{{Cormon}}},
  \bibinfo{author}{\bibfnamefont{M.}~\bibnamefont{{Fechner}}},
  \bibinfo{author}{\bibfnamefont{L.}~\bibnamefont{{Giot}}},
  \bibinfo{author}{\bibfnamefont{T.}~\bibnamefont{{Lasserre}}},
  \bibinfo{author}{\bibfnamefont{J.}~\bibnamefont{{Martino}}},
  \bibinfo{author}{\bibfnamefont{G.}~\bibnamefont{{Mention}}},
  \bibnamefont{et~al.}, \bibinfo{journal}{\prc} \textbf{\bibinfo{volume}{83}},
  \bibinfo{eid}{054615} (\bibinfo{year}{2011}), \eprint{1101.2663}.

\bibitem[{\citenamefont{{Mention} et~al.}(2011)\citenamefont{{Mention},
  {Fechner}, {Lasserre}, {Mueller}, {Lhuillier}, {Cribier}, and
  {Letourneau}}}]{2011PhRvD..83g3006M}
\bibinfo{author}{\bibfnamefont{G.}~\bibnamefont{{Mention}}},
  \bibinfo{author}{\bibfnamefont{M.}~\bibnamefont{{Fechner}}},
  \bibinfo{author}{\bibfnamefont{T.}~\bibnamefont{{Lasserre}}},
  \bibinfo{author}{\bibfnamefont{T.~A.} \bibnamefont{{Mueller}}},
  \bibinfo{author}{\bibfnamefont{D.}~\bibnamefont{{Lhuillier}}},
  \bibinfo{author}{\bibfnamefont{M.}~\bibnamefont{{Cribier}}},
  \bibnamefont{and}
  \bibinfo{author}{\bibfnamefont{A.}~\bibnamefont{{Letourneau}}},
  \bibinfo{journal}{\prd} \textbf{\bibinfo{volume}{83}}, \bibinfo{eid}{073006}
  (\bibinfo{year}{2011}), \eprint{1101.2755}.

\bibitem[{\citenamefont{{Huber}}(2011)}]{2011PhRvC..84b4617H}
\bibinfo{author}{\bibfnamefont{P.}~\bibnamefont{{Huber}}},
  \bibinfo{journal}{\prc} \textbf{\bibinfo{volume}{84}}, \bibinfo{eid}{024617}
  (\bibinfo{year}{2011}), \eprint{1106.0687}.

\bibitem[{\citenamefont{{Bahcall} et~al.}(1995)\citenamefont{{Bahcall},
  {Krastev}, and {Lisi}}}]{1995PhLB..348..121B}
\bibinfo{author}{\bibfnamefont{J.~N.} \bibnamefont{{Bahcall}}},
  \bibinfo{author}{\bibfnamefont{P.~I.} \bibnamefont{{Krastev}}},
  \bibnamefont{and} \bibinfo{author}{\bibfnamefont{E.}~\bibnamefont{{Lisi}}},
  \bibinfo{journal}{Physics Letters B} \textbf{\bibinfo{volume}{348}},
  \bibinfo{pages}{121} (\bibinfo{year}{1995}), \eprint{hep-ph/9411414}.

\bibitem[{\citenamefont{{Giunti} and {Laveder}}(2007)}]{2007MPLA...22.2499G}
\bibinfo{author}{\bibfnamefont{C.}~\bibnamefont{{Giunti}}} \bibnamefont{and}
  \bibinfo{author}{\bibfnamefont{M.}~\bibnamefont{{Laveder}}},
  \bibinfo{journal}{Modern Physics Letters A} \textbf{\bibinfo{volume}{22}},
  \bibinfo{pages}{2499} (\bibinfo{year}{2007}), \eprint{hep-ph/0610352}.

\bibitem[{\citenamefont{{Giunti} and {Laveder}}(2010)}]{2010PhRvD..82e3005G}
\bibinfo{author}{\bibfnamefont{C.}~\bibnamefont{{Giunti}}} \bibnamefont{and}
  \bibinfo{author}{\bibfnamefont{M.}~\bibnamefont{{Laveder}}},
  \bibinfo{journal}{\prd} \textbf{\bibinfo{volume}{82}}, \bibinfo{eid}{053005}
  (\bibinfo{year}{2010}), \eprint{1005.4599}.

\bibitem[{\citenamefont{{Giunti} and {Laveder}}(2011)}]{2011PhRvC..83f5504G}
\bibinfo{author}{\bibfnamefont{C.}~\bibnamefont{{Giunti}}} \bibnamefont{and}
  \bibinfo{author}{\bibfnamefont{M.}~\bibnamefont{{Laveder}}},
  \bibinfo{journal}{\prc} \textbf{\bibinfo{volume}{83}}, \bibinfo{eid}{065504}
  (\bibinfo{year}{2011}), \eprint{1006.3244}.

\bibitem[{\citenamefont{Dewhurst}(2015)}]{Dewhurst:2015aba}
\bibinfo{author}{\bibfnamefont{D.}~\bibnamefont{Dewhurst}}
  (\bibinfo{collaboration}{T2K}), in \emph{\bibinfo{booktitle}{{Topical
  Research Meeting on Prospects in Neutrino Physics (NuPhys2014) London, UK,
  December 15-17, 2014}}} (\bibinfo{year}{2015}), \eprint{1504.08237},
  \urlprefix\url{https://inspirehep.net/record/1365546/files/arXiv:1504.08237.pdf}.

\bibitem[{\citenamefont{An et~al.}(2014)}]{An:2014bik}
\bibinfo{author}{\bibfnamefont{F.~P.} \bibnamefont{An}} \bibnamefont{et~al.}
  (\bibinfo{collaboration}{Daya Bay}), \bibinfo{journal}{Phys. Rev. Lett.}
  \textbf{\bibinfo{volume}{113}}, \bibinfo{pages}{141802}
  (\bibinfo{year}{2014}), \eprint{1407.7259}.

\bibitem[{\citenamefont{Acciarri et~al.}(2015)}]{Acciarri:2015uup}
\bibinfo{author}{\bibfnamefont{R.}~\bibnamefont{Acciarri}} \bibnamefont{et~al.}
  (\bibinfo{collaboration}{DUNE}) (\bibinfo{year}{2015}), \eprint{1512.06148}.

\bibitem[{\citenamefont{Acciarri
  et~al.}(2016{\natexlab{a}})}]{Acciarri:2016crz}
\bibinfo{author}{\bibfnamefont{R.}~\bibnamefont{Acciarri}} \bibnamefont{et~al.}
  (\bibinfo{collaboration}{DUNE}) (\bibinfo{year}{2016}{\natexlab{a}}),
  \eprint{1601.05471}.

\bibitem[{\citenamefont{Strait et~al.}(2016)}]{Strait:2016mof}
\bibinfo{author}{\bibfnamefont{J.}~\bibnamefont{Strait}} \bibnamefont{et~al.}
  (\bibinfo{collaboration}{DUNE}) (\bibinfo{year}{2016}), \eprint{1601.05823}.

\bibitem[{\citenamefont{Acciarri
  et~al.}(2016{\natexlab{b}})}]{Acciarri:2016ooe}
\bibinfo{author}{\bibfnamefont{R.}~\bibnamefont{Acciarri}} \bibnamefont{et~al.}
  (\bibinfo{collaboration}{DUNE}) (\bibinfo{year}{2016}{\natexlab{b}}),
  \eprint{1601.02984}.

\bibitem[{\citenamefont{Hollander and Mocioiu}(2015)}]{Hollander:2014iha}
\bibinfo{author}{\bibfnamefont{D.}~\bibnamefont{Hollander}} \bibnamefont{and}
  \bibinfo{author}{\bibfnamefont{I.}~\bibnamefont{Mocioiu}},
  \bibinfo{journal}{Phys. Rev.} \textbf{\bibinfo{volume}{D91}},
  \bibinfo{pages}{013002} (\bibinfo{year}{2015}), \eprint{1408.1749}.

\bibitem[{\citenamefont{{Masud} and {Mehta}}(2016)}]{2016arXiv160301380M}
\bibinfo{author}{\bibfnamefont{M.}~\bibnamefont{{Masud}}} \bibnamefont{and}
  \bibinfo{author}{\bibfnamefont{P.}~\bibnamefont{{Mehta}}},
  \bibinfo{journal}{ArXiv e-prints}  (\bibinfo{year}{2016}),
  \eprint{1603.01380}.

\bibitem[{\citenamefont{{de Gouv{\^e}a} and
  {Kelly}}(2015)}]{2015arXiv151105562D}
\bibinfo{author}{\bibfnamefont{A.}~\bibnamefont{{de Gouv{\^e}a}}}
  \bibnamefont{and} \bibinfo{author}{\bibfnamefont{K.~J.}
  \bibnamefont{{Kelly}}}, \bibinfo{journal}{ArXiv e-prints}
  (\bibinfo{year}{2015}), \eprint{1511.05562}.

\bibitem[{\citenamefont{{Coloma}}(2016)}]{2016JHEP...03..016C}
\bibinfo{author}{\bibfnamefont{P.}~\bibnamefont{{Coloma}}},
  \bibinfo{journal}{Journal of High Energy Physics}
  \textbf{\bibinfo{volume}{3}}, \bibinfo{eid}{16} (\bibinfo{year}{2016}),
  \eprint{1511.06357}.

\bibitem[{\citenamefont{{Gandhi} et~al.}(2015)\citenamefont{{Gandhi}, {Kayser},
  {Masud}, and {Prakash}}}]{2015JHEP...11..039G}
\bibinfo{author}{\bibfnamefont{R.}~\bibnamefont{{Gandhi}}},
  \bibinfo{author}{\bibfnamefont{B.}~\bibnamefont{{Kayser}}},
  \bibinfo{author}{\bibfnamefont{M.}~\bibnamefont{{Masud}}}, \bibnamefont{and}
  \bibinfo{author}{\bibfnamefont{S.}~\bibnamefont{{Prakash}}},
  \bibinfo{journal}{Journal of High Energy Physics}
  \textbf{\bibinfo{volume}{11}}, \bibinfo{eid}{39} (\bibinfo{year}{2015}),
  \eprint{1508.06275}.

\bibitem[{\citenamefont{{Agarwalla} et~al.}(2016)\citenamefont{{Agarwalla},
  {Sachi Chatterjee}, and {Palazzo}}}]{2016arXiv160303759A}
\bibinfo{author}{\bibfnamefont{S.~K.} \bibnamefont{{Agarwalla}}},
  \bibinfo{author}{\bibfnamefont{S.}~\bibnamefont{{Sachi Chatterjee}}},
  \bibnamefont{and}
  \bibinfo{author}{\bibfnamefont{A.}~\bibnamefont{{Palazzo}}},
  \bibinfo{journal}{ArXiv e-prints}  (\bibinfo{year}{2016}),
  \eprint{1603.03759}.

\bibitem[{\citenamefont{{Berryman} et~al.}(2015)\citenamefont{{Berryman}, {de
  Gouvea}, {Kelly}, and {Kobach}}}]{2015arXiv150703986B}
\bibinfo{author}{\bibfnamefont{J.~M.} \bibnamefont{{Berryman}}},
  \bibinfo{author}{\bibfnamefont{A.}~\bibnamefont{{de Gouvea}}},
  \bibinfo{author}{\bibfnamefont{K.~J.} \bibnamefont{{Kelly}}},
  \bibnamefont{and} \bibinfo{author}{\bibfnamefont{A.}~\bibnamefont{{Kobach}}},
  \bibinfo{journal}{ArXiv e-prints}  (\bibinfo{year}{2015}),
  \eprint{1507.03986}.

\bibitem[{\citenamefont{{Berryman} et~al.}(2016)\citenamefont{{Berryman}, {de
  Gouv{\^e}a}, {Kelly}, {Peres}, and {Tabrizi}}}]{2016arXiv160300018B}
\bibinfo{author}{\bibfnamefont{J.~M.} \bibnamefont{{Berryman}}},
  \bibinfo{author}{\bibfnamefont{A.}~\bibnamefont{{de Gouv{\^e}a}}},
  \bibinfo{author}{\bibfnamefont{K.~J.} \bibnamefont{{Kelly}}},
  \bibinfo{author}{\bibfnamefont{O.~L.~G.} \bibnamefont{{Peres}}},
  \bibnamefont{and}
  \bibinfo{author}{\bibfnamefont{Z.}~\bibnamefont{{Tabrizi}}},
  \bibinfo{journal}{ArXiv e-prints}  (\bibinfo{year}{2016}),
  \eprint{1603.00018}.

\bibitem[{\citenamefont{Bhattacharya et~al.}(2012)\citenamefont{Bhattacharya,
  Thalapillil, and Wagner}}]{Bhattacharya:2011ee}
\bibinfo{author}{\bibfnamefont{B.}~\bibnamefont{Bhattacharya}},
  \bibinfo{author}{\bibfnamefont{A.~M.} \bibnamefont{Thalapillil}},
  \bibnamefont{and} \bibinfo{author}{\bibfnamefont{C.~E.~M.}
  \bibnamefont{Wagner}}, \bibinfo{journal}{Phys. Rev.}
  \textbf{\bibinfo{volume}{D85}}, \bibinfo{pages}{073004}
  (\bibinfo{year}{2012}), \eprint{1111.4225}.

\bibitem[{\citenamefont{{Klop} and {Palazzo}}(2015)}]{2015PhRvD..91g3017K}
\bibinfo{author}{\bibfnamefont{N.}~\bibnamefont{{Klop}}} \bibnamefont{and}
  \bibinfo{author}{\bibfnamefont{A.}~\bibnamefont{{Palazzo}}},
  \bibinfo{journal}{\prd} \textbf{\bibinfo{volume}{91}}, \bibinfo{eid}{073017}
  (\bibinfo{year}{2015}), \eprint{1412.7524}.

\bibitem[{\citenamefont{{Palazzo}}(2015)}]{2015arXiv150903148P}
\bibinfo{author}{\bibfnamefont{A.}~\bibnamefont{{Palazzo}}},
  \bibinfo{journal}{ArXiv e-prints}  (\bibinfo{year}{2015}),
  \eprint{1509.03148}.

\bibitem[{\citenamefont{{Blennow} et~al.}(2014)\citenamefont{{Blennow},
  {Coloma}, and {Fernandez-Martinez}}}]{2014JHEP...12..120B}
\bibinfo{author}{\bibfnamefont{M.}~\bibnamefont{{Blennow}}},
  \bibinfo{author}{\bibfnamefont{P.}~\bibnamefont{{Coloma}}}, \bibnamefont{and}
  \bibinfo{author}{\bibfnamefont{E.}~\bibnamefont{{Fernandez-Martinez}}},
  \bibinfo{journal}{Journal of High Energy Physics}
  \textbf{\bibinfo{volume}{12}}, \bibinfo{eid}{120} (\bibinfo{year}{2014}),
  \eprint{1407.1317}.

\bibitem[{\citenamefont{Choudhary et~al.}(2012)\citenamefont{Choudhary, Gandhi,
  R., Mishra, and Strait}}]{near}
\bibinfo{author}{\bibfnamefont{B.}~\bibnamefont{Choudhary}},
  \bibinfo{author}{\bibfnamefont{R.}~\bibnamefont{Gandhi}},
  \bibinfo{author}{\bibfnamefont{M.~S.} \bibnamefont{R.}},
  \bibinfo{author}{\bibnamefont{Mishra}}, \bibnamefont{and}
  \bibinfo{author}{\bibfnamefont{J.}~\bibnamefont{Strait}}
  (\bibinfo{year}{2012}),
  \urlprefix\url{http://lbne2-docdb.fnal.gov/cgi-bin/RetrieveFile?docid=6704&filename=LBNE-India-DPR-V12-Science.pdf&version=1}.

\bibitem[{\citenamefont{{Huber} et~al.}(2005)\citenamefont{{Huber}, {Lindner},
  and {Winter}}}]{2005CoPhC.167..195H}
\bibinfo{author}{\bibfnamefont{P.}~\bibnamefont{{Huber}}},
  \bibinfo{author}{\bibfnamefont{M.}~\bibnamefont{{Lindner}}},
  \bibnamefont{and} \bibinfo{author}{\bibfnamefont{W.}~\bibnamefont{{Winter}}},
  \bibinfo{journal}{Computer Physics Communications}
  \textbf{\bibinfo{volume}{167}}, \bibinfo{pages}{195} (\bibinfo{year}{2005}),
  \eprint{hep-ph/0407333}.

\bibitem[{\citenamefont{{Huber} et~al.}(2007)\citenamefont{{Huber}, {Kopp},
  {Lindner}, {Rolinec}, and {Winter}}}]{2007CoPhC.177..432H}
\bibinfo{author}{\bibfnamefont{P.}~\bibnamefont{{Huber}}},
  \bibinfo{author}{\bibfnamefont{J.}~\bibnamefont{{Kopp}}},
  \bibinfo{author}{\bibfnamefont{M.}~\bibnamefont{{Lindner}}},
  \bibinfo{author}{\bibfnamefont{M.}~\bibnamefont{{Rolinec}}},
  \bibnamefont{and} \bibinfo{author}{\bibfnamefont{W.}~\bibnamefont{{Winter}}},
  \bibinfo{journal}{Computer Physics Communications}
  \textbf{\bibinfo{volume}{177}}, \bibinfo{pages}{432} (\bibinfo{year}{2007}),
  \eprint{hep-ph/0701187}.

\bibitem[{\citenamefont{Cavanna}(2016)}]{Cavanna:2144868}
\bibinfo{author}{\bibfnamefont{F.}~\bibnamefont{Cavanna}}
  (\bibinfo{collaboration}{NP04 Collaboration}), \bibinfo{type}{Tech. Rep.}
  \bibinfo{number}{CERN-SPSC-2016-018. SPSC-SR-185},
  \bibinfo{institution}{CERN}, \bibinfo{address}{Geneva}
  (\bibinfo{year}{2016}), \urlprefix\url{https://cds.cern.ch/record/2144868}.

\bibitem[{\citenamefont{{LBNE Collaboration} et~al.}(2013)\citenamefont{{LBNE
  Collaboration}, {Adams}, {Adams}, {Akiri}, {Alion}, {Anderson},
  {Andreopoulos}, {Andrews}, {Anghel}, {Costa dos Anjos}
  et~al.}}]{2013arXiv1307.7335L}
\bibinfo{author}{\bibnamefont{{LBNE Collaboration}}},
  \bibinfo{author}{\bibfnamefont{C.}~\bibnamefont{{Adams}}},
  \bibinfo{author}{\bibfnamefont{D.}~\bibnamefont{{Adams}}},
  \bibinfo{author}{\bibfnamefont{T.}~\bibnamefont{{Akiri}}},
  \bibinfo{author}{\bibfnamefont{T.}~\bibnamefont{{Alion}}},
  \bibinfo{author}{\bibfnamefont{K.}~\bibnamefont{{Anderson}}},
  \bibinfo{author}{\bibfnamefont{C.}~\bibnamefont{{Andreopoulos}}},
  \bibinfo{author}{\bibfnamefont{M.}~\bibnamefont{{Andrews}}},
  \bibinfo{author}{\bibfnamefont{I.}~\bibnamefont{{Anghel}}},
  \bibinfo{author}{\bibfnamefont{J.~C.} \bibnamefont{{Costa dos Anjos}}},
  \bibnamefont{et~al.}, \bibinfo{journal}{ArXiv e-prints}
  (\bibinfo{year}{2013}), \eprint{1307.7335}.

\bibitem[{\citenamefont{Bishai}(Private communication)}]{marie}
\bibinfo{author}{\bibfnamefont{M.}~\bibnamefont{Bishai}}
  (\bibinfo{year}{Private communication}).

\bibitem[{\citenamefont{Kopp}(2008)}]{Kopp:2006wp}
\bibinfo{author}{\bibfnamefont{J.}~\bibnamefont{Kopp}}, \bibinfo{journal}{Int.
  J. Mod. Phys.} \textbf{\bibinfo{volume}{C19}}, \bibinfo{pages}{523}
  (\bibinfo{year}{2008}), \bibinfo{note}{erratum ibid.\ {\bf C19} (2008) 845},
  \eprint{physics/0610206}.

\bibitem[{\citenamefont{Kopp et~al.}(2008)\citenamefont{Kopp, Lindner, Ota, and
  Sato}}]{Kopp:2007ne}
\bibinfo{author}{\bibfnamefont{J.}~\bibnamefont{Kopp}},
  \bibinfo{author}{\bibfnamefont{M.}~\bibnamefont{Lindner}},
  \bibinfo{author}{\bibfnamefont{T.}~\bibnamefont{Ota}}, \bibnamefont{and}
  \bibinfo{author}{\bibfnamefont{J.}~\bibnamefont{Sato}},
  \bibinfo{journal}{Phys. Rev.} \textbf{\bibinfo{volume}{D77}},
  \bibinfo{pages}{013007} (\bibinfo{year}{2008}), \eprint{0708.0152}.

\bibitem[{\citenamefont{Antonello et~al.}(2015)}]{Antonello:2015lea}
\bibinfo{author}{\bibfnamefont{M.}~\bibnamefont{Antonello}}
  \bibnamefont{et~al.} (\bibinfo{collaboration}{LAr1-ND, ICARUS-WA104,
  MicroBooNE}) (\bibinfo{year}{2015}), \eprint{1503.01520}.

\bibitem[{Osc(2013)}]{OscSNS:2013hua}
 (\bibinfo{year}{2013}), \eprint{1305.4189}.

\bibitem[{\citenamefont{Agarwalla and Huber}(2011)}]{Agarwalla:2010zu}
\bibinfo{author}{\bibfnamefont{S.~K.} \bibnamefont{Agarwalla}}
  \bibnamefont{and} \bibinfo{author}{\bibfnamefont{P.}~\bibnamefont{Huber}},
  \bibinfo{journal}{Phys. Lett.} \textbf{\bibinfo{volume}{B696}},
  \bibinfo{pages}{359} (\bibinfo{year}{2011}), \eprint{1007.3228}.

\bibitem[{\citenamefont{Spitz}(2012)}]{Spitz:2012gp}
\bibinfo{author}{\bibfnamefont{J.}~\bibnamefont{Spitz}},
  \bibinfo{journal}{Phys. Rev.} \textbf{\bibinfo{volume}{D85}},
  \bibinfo{pages}{093020} (\bibinfo{year}{2012}), \eprint{1203.6050}.

\bibitem[{\citenamefont{Antonello et~al.}(2012)}]{Antonello:2012hf}
\bibinfo{author}{\bibfnamefont{M.}~\bibnamefont{Antonello}}
  \bibnamefont{et~al.} (\bibinfo{year}{2012}), \eprint{1203.3432}.

\bibitem[{\citenamefont{Jones}(2013)}]{Jones:2011ci}
\bibinfo{author}{\bibfnamefont{B.~J.~P.} \bibnamefont{Jones}},
  \bibinfo{journal}{J. Phys. Conf. Ser.} \textbf{\bibinfo{volume}{408}},
  \bibinfo{pages}{012028} (\bibinfo{year}{2013}), \eprint{1110.1678}.

\bibitem[{\citenamefont{Koshkarev}(1974)}]{Koshkarev:1974my}
\bibinfo{author}{\bibfnamefont{D.~G.} \bibnamefont{Koshkarev}}
  (\bibinfo{year}{1974}).

\bibitem[{\citenamefont{Neuffer}(1980)}]{Neuffer:1980ru}
\bibinfo{author}{\bibfnamefont{D.}~\bibnamefont{Neuffer}},
  \bibinfo{journal}{eConf} \textbf{\bibinfo{volume}{C801002}},
  \bibinfo{pages}{199} (\bibinfo{year}{1980}).

\end{thebibliography}

\end{document}